\documentclass[a4paper,fleqn,usenatbib]{mnras}

\usepackage{newtxtext,newtxmath}
\usepackage[T1]{fontenc}
\usepackage{ae,aecompl}

\usepackage{graphicx}	
\usepackage{amsmath}	
\usepackage{amssymb}	

\newcommand{\ML}{$M/L$}
\newcommand{\MLref}{$\Upsilon_{\rm ref}$}  
\newcommand{\MLstell}{$\Upsilon$}
\newcommand{\rein}{$R_{\rm Ein}$}
\newcommand{\reff}{$R_{\rm eff}$}

\title[The MUSE lens search]{MNELLS: The MUSE Nearby Early-Type Galaxy Lens Locator Survey}

\author[Collier W. et al.]{
William P. Collier,\thanks{E-mail: william.p.collier@durham.ac.uk}
Russell J. Smith,
John R. Lucey
\\
Centre for Extragalactic Astronomy, Departments of Physics, University of Durham, Durham DH1 3LE, UK\\
}

\date{MNRAS submitted}

\pubyear{2019}

\begin{document}
\label{firstpage}
\pagerange{\pageref{firstpage}--\pageref{lastpage}}
\maketitle

\begin{abstract}

Low-redshift strong-lensing galaxies can provide robust measurements of the stellar mass-to-light ratios in early-type galaxies (ETG), and hence constrain variations in the stellar initial mass function (IMF). At present, only a few such systems are known. Here, we report the first results from a blind search for gravitationally-lensed emission line sources behind 52 massive $z$\,$<$\,0.07 ETGs with MUSE integral field spectroscopy. For 16 galaxies, new observations were acquired, whilst the other 36 were analysed from archival data. This project has previously yielded one confirmed galaxy-scale strong lens (J0403-0239) which we report in an earlier paper. J0403-0239 has since received follow-up observations, presented here, which indicate support for our earlier IMF results. Three cluster-scale, and hence dark-matter-dominated, lensing systems were also discovered (central galaxies of A4059, A2052 and AS555).  For nine further galaxies, we detect a singly-imaged but closely-projected source within 6\,arcsec (including one candidate with sources at three different redshifts); such cases can be exploited to derive upper limits on the IMF mass-excess factor, $\alpha$. Combining the new lens and new upper limits, with the previously-discovered systems, we infer an average $\langle \alpha \rangle$\,=\,1.06\,$\pm$\,0.08 (marginalised over the intrinsic scatter), which is  inconsistent with a Salpeter-like IMF ($\alpha$\,=\,1.55) at the 6$\sigma$ level. We test the detection threshold in these short-exposure MUSE observations with the injection and recovery of simulated sources, and predict that one in twenty-five observations is expected to yield a new strong-lens system. Our observational results are consistent with this expected yield.



\end{abstract}


\begin{keywords}
gravitational lensing: strong --- galaxies: elliptical and lenticular, cD --- galaxies:stellar content
\end{keywords}



\section{Introduction}
\label{sec:Intro}

\citet{Zwicky1937} first described how a nearby galaxy could act as a gravitational lens to a more distant source, and the resulting lensed images could be of sufficient size and surface brightness to be observable. The first discovered lensed images were a pair of quasars at $z$\,=\,1.41 separated by 6\,arcsec with near-identical spectra \citep{Walsh1979}. Recent lensing studies, with better characterised lens light profiles, provide key constraints on a wide range of astrophysical quantities, such as H$_{\rm 0}$, general relativity and the nature of dark matter \citep[e.g.][]{Wong2019,Collett2018,Ritondale2019}.



Strong lensing measures the total mass, including dark matter (DM) as well as stars, projected within the Einstein radius, ($R_{\rm Ein}$). In cases where the relative contributions of the two components can be distinguished, the stellar mass-to-light ratio ($\Upsilon$) can be determined. This is particularly important when investigating the stellar initial mass function (IMF), which is a key component of galaxy evolution, and for interpreting observed properties e.g. estimating stellar masses. The IMF is probed via the mass-excess parameter ($\alpha$);


\begin{equation}
    \alpha = \dfrac{\Upsilon}{\Upsilon_{\rm Ref}}
\end{equation}

which compares a measured \MLstell\ to a reference mass-to-light ratio (\MLref) for a modelled stellar populations with a fixed IMF. Disentangling the dark and stellar matter, in general, requires additional information which can in principle be determined from stellar dynamics \citep[i.e. SLACS,][]{Treu2010b}. However this technique requires further modelling assumptions, and appears to require the addition of \MLstell\ gradients to match similar measurements from weak lensing constraints \citep{Sonnenfeld2018}. Within a galaxy, the stars are more centrally concentrated than the DM halo, e.g. a NFW halo \citep*{NFW1996}. Lenses in which the \rein\ occurs at a fraction of the effective radius (\reff) directly probes the stellar content of a galaxy. For low redshift lenses ($z$\,$\leq$\,0.1), the required critical density for forming multiple images is higher, (compared to the same lens, but more distant from the observer), and is therefore exceeded only at small physical radius where the stars contribute up to 80 percent of the mass.


Recent efforts to detect low redshift strong lensing early-type galaxies (ETGs) have been driven largely by the seeming discrepancy between the measured $\alpha$ parameters of the SLACS lenses ($\langle z \rangle$\,$\sim$\,0.2), compared to those from the SINFONI nearby elliptical lens locator survey \citep[SNELLS,][$\langle z \rangle$\,$\sim$\,0.03]{SLC2015}. These two samples measure significantly different IMFs within massive (high velocity dispersion) ETGs i.e. at the 3$\sigma$ level. Additionally, these SNELLS lenses are valuable as they provide a `golden sample' for which all three of the main IMF tracing techniques, strong-lensing, stellar dynamics and spectral synthesis, can be applied, and therefore contrasted \citep{Newman2017}. However observable strong-gravitational lensing by nearby massive ETGs is inherently rare.



Techniques used to discover the current sample of galaxy-scale lenses fall broadly into two categories; those searching spatially within imaging, and those using single-fibre spectroscopy to detect background emission line objects. Within imaging the preferred methods focus on identifying either elongated arcs \citep[i.e. strong lensing challenge,][]{Metcalf2018}, or multiple sources with similar colour properties distinct from the foreground galaxies \citep{Lemon2018,Lemon2019,Lucey2018,Delchambre2019,Rusu2019}. The second category yields lenses from spectral features distinct from the foreground galaxy in redshift space, within large spectroscopic surveys (e.g. SDSS). In most systems, the background source is detected in emission \citep[e.g. SLACS, BELLS, BELLS--GALLERY,][]{Bolton2006,Brownstein2012}. However, it is also possible to detect background absorption line systems \citep[i.e. Early-Early Lenses,][]{Oldham2017}. The foreground lens galaxy is fit with a model spectrum, and then the residual is searched for higher redshift residual features. Due to the small angular size of a fibre (e.g. 3\,arcsec, SDSS), if a background source is detected, it is likely to lie within the multiply-imaged region of the source plane.




For lens systems identified through either of the classical techniques, the discovery data are generally not sufficient for a full analysis. Imaging only reveals the number, and fluxes of potential images, but does not provide the redshifts which are required to confirm a single source of origin. Conversely spectroscopy only confirms the redshift of a background object, but not the configuration. Therefore neither of these are observationally efficient. Furthermore the discovered lens systems are biased by the detection method. To detect within imaging alone, the lensed sources must be resolved against the foreground lens light, and hence frequently lie at high angular separation (compared to the lens \reff). Spectroscopic searches however, are limited by the fibre size. To detect an emission line enough of the lensed image flux must be contained within a fibre, which may not be the case for highly asymmetric systems. Therefore both of these techniques may fail to detect valid strong-lensing systems.


The advent of large field-of-view (FoV) integral field units (IFU) with sufficient angular and spectral resolution, offers a third approach for lens discovery. IFU observations make use of spectral pixels, to uncover the background emission lines, along with the spatial resolution to simultaneously search for redshift matched, spatially separated images. Due to the large FoV, the position of lensed images are no longer limited to the small fibre aperture size regime. With the spectral resolution, fainter background sources may be discovered at the peak contrast between the lensed emission, and the foreground continuum. High resolution imaging can be acquired for more detailed mass modelling, however measurements from poorer resolution IFU data has been found to match closely with {\it Hubble Space Telescope} (HST) imaging \citep*{Collier2018a}.

To date, almost all low redshift lenses have been discovered with IFU observations. Two lenses, SNL-1, and SNL-2 were the subject of targeted observations with the SINFONI instrument for the SNELLS survey \citep*{SLC2015}. A further lens, J0403-5055 \citep*{Collier2018b}, was discovered within archival data from the Multi-Unit Spectroscopic Explorer \citep[MUSE,][]{Bacon2014}, with the observation taken for alternative (non-lensing) science. This lens was independently reported by \citet{Galbany2018}. With the completion of current large IFU surveys, \citep[i.e SDSS-IV MaNGA, SAMI,][]{Bundy2015,Bryant2015}, of order 10$^{4}$ ETGs will be observed, and searched for lensed images. Future surveys such as HECTOR \citep{BlandHawthorn2015} will expand this sample towards 10$^{5}$ galaxies. The MaNGA survey targets luminous galaxies with a low median redshift ($z$\,$\sim$\,0.05), and with the DR14 data-release \citet{Smith2017} discovered one new lens, and new candidate systems were reported from a sample of 2812 galaxies by \citet{Talbot2018}. 

A key difference between the SNELLS approach and the MaNGA and SAMI surveys, is the selection criteria for the targeted galaxies. SNELLS selected only the most massive ETGs as they have the largest lensing cross-sections. In doing so, the number of galaxies which must be observed to return a significant yield of lenses is greatly reduced. Although SNELLS discovered two lenses, selecting to use SINFONI for this search has limitations. The FoV of 8\,$\times$\,8\,arcsec, leads to highly asymmetric systems still being contaminated by FoV edge effects, see SNL-2, \citep{SLC2015}. In addition the wavelength range 1.1--2.45\,$\mu$m limits the number of emitters being probed, as only sources with redshift greater than $\sim$\,0.7 will have strong emission lines, such as H\,$\alpha$, in the detection range. Furthermore there is a significant sky background incurred by working in the near-IR. 

To overcome some of the limitations, we extend the technique to a wide FoV optical IFU. The MUSE Nearby Early-Type Galaxy Lens Locator Survey (MNELLS) utilises the 1\,$\times$\,1 arcmin FoV to detect even the most asymmetric systems, whilst retaining a high angular resolution (0.2 sq. arcsec). The wavelength range of 4750\,--\,9300\,\AA\ probes [{\sc O\,ii}] emitters up to a redshift of $z$\,=\,1.5, and Ly-$\alpha$ above $z$\,$\sim$\,3. Here we work with data from the ESO Period 101 observations (PI: Smith). As the MUSE IFU has been in operation since 2014, we also select galaxies from the large public archive.



In this paper we report the results from our targeted and archival lens searches. In Section \ref{sec:Sample} we present the sample selection for the targetted observations and those used to pick from the archive. In Section \ref{sec:gseld} we present the process for identifying the emission-line sources within the datacubes. We report detections and promising candidates from the sample in Section \ref{sec:ICL}. Our lensing analysis is in Section \ref{sec:LA}, and we asses the detection limits of our search in Section \ref{sec:RoD}. In Section \ref{sec:Disc} we compare our reliability and detection limits within our observations compared to those expected and comment on implications for future searches.

In this paper we adopt cosmology from \citet{Planck2018}, i.e. H$_0$ = 67.4 km s$^{-1}$ Mpc$^{-1}$, $\Omega _{\rm m}$ = 0.315 and $\Omega_{\Lambda}$ = 0.685. 

\section{Data}
\label{sec:Sample}

In this Section we outline the sample selection for the galaxies which will be used in our analysis. We describe our MUSE targeted programme in Section \ref{sec:MNELLS}, and then our selection of archival MUSE observations in Section \ref{sec:MUSEarch}.




\subsection{Targeted Sample}
\label{sec:MNELLS}

Our MUSE survey targets massive ETGs, selected by velocity dispersion, in the local universe. These massive galaxies maximise the lensing cross-section per target, and hence increase the probability of discovering a lens. In this subsection we outline our target selection criteria. 



\subsubsection{Target Selection}
\label{sec:MNELLS_ts}

MNELLS builds upon the previous SNELLS survey, and we used the following selection criteria, similar to the earlier work:

1. A redshift, $z$\,$<$\,0.060.

2. A stellar velocity dispersion measured from high S/N spectra, from either the 6dFGSv \citep[][]{Campbell2014}, or SDSS \citep[][]{York2000}, ($\sigma_{\rm 6dF}$\,$>$\,300\,km\,s$^{-1}$, $\sigma_{\rm SDSS}$\,$>$\,310\,km\,s$^{-1}$, to allow for the differing fibre sizes). 

3. The galaxy must not lie in a rich cluster/massive group environment. This prevents additional complexity in the modelling to account for either the cluster potential, or lensing effects (external shear) from other nearby massive galaxies. We judge this using the NASA/IPAC Extragalactic Database.

4. The galaxy must be observable in ESO Period 101, April\,-\,September from the VLT.  The targets were chosen to scatter across the full Right Ascension range available for the semester, with a preference for southern Declinations, to take advantage of the wind restrictions on northern pointings at Paranal.


\subsubsection{Observations and Data Reduction}
\label{sec:MNELLS_odr}

The MUSE observations were undertaken in service mode (April\,-\,September 2018). Each galaxy was assigned two 40min observing blocks (OB, rotated by 45 deg), composed of four 380\,second exposures. Using wide-field no-AO mode, each frame consists of a 1$\times$1 arcmin$^2$ FoV, with a pixel scale of 0.2\,$\times$\,0.2\,arcsec$^2$ and a wavelength resolution of $\sim$\,2.7\,\AA, sampled at 1.25\,\AA/pix.   

A total of 16 candidates were observed, and of these, 14 have both OBs (i.e. full depth). The galaxies with at least one OB are shown in Table \ref{tab:MNELLS}. Each observation was retrieved as a pipeline-reduced file as provided by ESO.

\subsection{MUSE Archival Sample}
\label{sec:MUSEarch}

There are over 9000 existing MUSE observations publicly available. Therefore, we supplement our targeted programme with archival data. As the data already exists we relax our criteria from the very restrictive targeted selection of Section \ref{sec:MNELLS}\footnote{In principle the archive could be searched for strong gravitational lenses independent of specific morphologies or redshift. However this is beyond the scope of this work, specific to the IMF of massive ETGs.}. Particularly, we allow BCGs into the sample\footnote{BCGs are excluded from the targeted survey as the lensing mass is typically dominated by DM, and the galaxy has a low surface brightness and hence mass density. This leads to an \rein\ significantly larger than found in a typical isolated galaxy. Hence, the DM modelling dominates the uncertainty on the constraints of the IMF.}  as these can act as strong-lenses for the same background emitters. In this subsection we outline our selection criteria for the supplementary sample drawn from the MUSE archive.





\subsubsection{Target Selection and observations}
\label{sec:arch_ts}

\begin{figure}
	\centering
	\includegraphics[width=\linewidth,trim={0 2cm 0 0}]{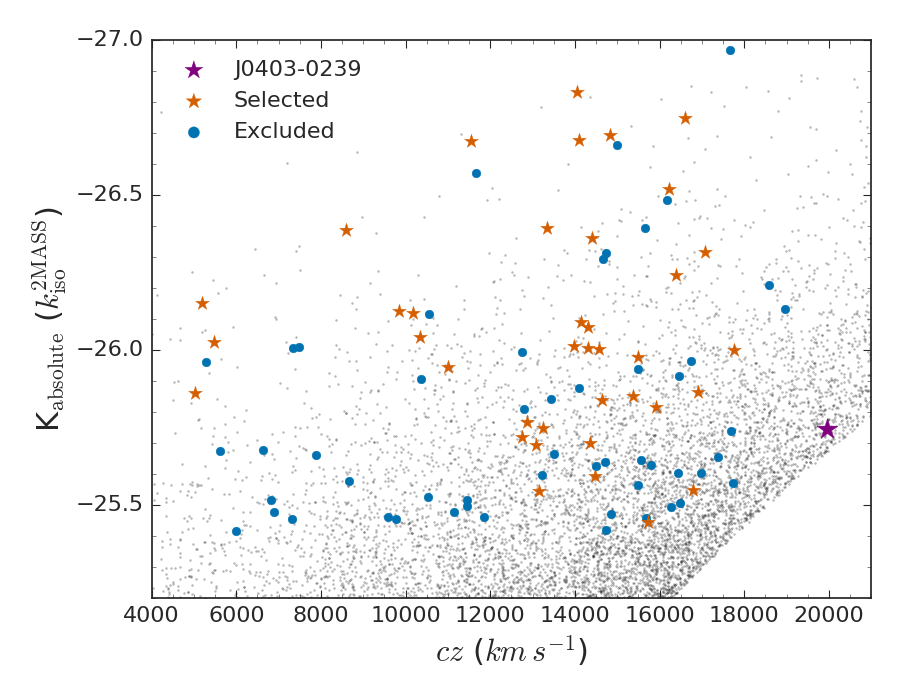}
	\caption{The redshift and absolute magnitude distribution of the 2MRS galaxies (background points), with those previously observed with MUSE (red, blue and purple points). Out of those matched to the MUSE archive we exclude a large number of systems (usually due to a complex light profile, morphology or companion galaxies), in blue, and mark the selected galaxies with red points. The confirmed lens J0403-0239 is shown in purple.}
	\label{fig:musesel}
\end{figure}

In order to build a sample of low-$z$ galaxies from the MUSE archive, we use positions, redshifts and luminosities from the 2MASS Redshift Survey \citep[2MRS,][]{Huchra2012}. We use the 2MRS due to its high completeness in our redshift range. We select galaxies with redshifts 0.01\,$\leq$\,$z$\,$\leq$\,0.07, and massive galaxies with a cut on the absolute K band magnitude (as a proxy for stellar mass), at K\,$\leq$\,--25.4 mag. We show the redshift distribution of the matched galaxies against their absolute magnitude in Figure \ref{fig:musesel}.



We visually inspect the MUSE collapsed datacube product, in order to exclude galaxies which would require complex lens analysis (i.e. very nearby similarly sized galaxies, or an irregular light profile). We also exclude galaxies which may have complex stellar populations (i.e. extended/strong emission, and mergers), or are of a spiral morphology. Our final archival sample has 36 galaxies.

We searched all observations which were publicly available as of February 2019. We list in Table \ref{tab:archive}, all of the investigated galaxy properties, along with the run ID and exposure times. Many of the observations were acquired by the `MUSE most massive galaxy (M3G)' survey \citep[][PI: Emsellem]{Krajnovic2018}.

For each galaxy in the sample we select the deepest available observation, which in many cases is the MUSE-DEEP data product. The datacubes are acquired as an ESO pipeline reduced final product. The selected observations range from exposure times comparable to, or shorter than our MNELLS sample, $<$\,3600 seconds, or much longer, $>$10000 seconds and have varied seeing condidtions.



\section{Galaxy subtraction and emission line detection }
\label{sec:gseld}


In this Section we outline our method to detect background emission line sources. In the SNELLS survey, each datacube was visually inspected to search for background emitters. Here, we automate the process, incorporating a search algorithm as a first pass. Then we visually inspect each detection to confirm the reliability. 

The detection process consists of two main steps: 1) the removal of the foreground lens candidate and other nearby bright continuum objects and 2) the detection and identification of emitters within the continuum subtracted datacubes. 


To help isolate compact narrow emission features, we first subtract a continuum spectrum from each pixel computed using a running median with a 75\,\AA\ bandwidth. The residuals from this step are next fitted with an elliptical profile computed for each wavelength channel independently, with the centre, ellipticity and position angle fixed to match the target galaxy. When other bright sources are present in the field, their residuals are modelled similarly. This process works well except near strong spectral features in fast-rotating galaxies. After subtracting these profiles, the remaining residuals are normalised to the local standard deviation (estimated using quantiles for robustness), and smoothed with a Gaussian of 0.6\,arcsec FWHM in the spatial directions and 3.5\,\AA\ FWHM in wavelength.

We developed a python-based framework for detecting emission line features within the filtered and smoothed datacubes using routines from the {\sc scipy.ndimages.label} package. This package groups connected pixels above a threshold, allowing the spatially and spectrally extended peaks, due to the emission line features, to be identified. The selection of the threshold involves a trade-off between the number of spurious detections to inspect manually, and those faint sources which may be missed.


The smoothed residual datacube is separated into 40\,\AA\ slices with an overlap of 5\AA\, and a labelling threshold is applied. For each labelled region, the number of pixels and the spatial extent are required to be above thresholds of 10 and 2\,$\times$\,2 respectively. Those which do not meet these criteria are not considered for further processing. These are usually artefacts from the subtraction of bright sky lines, or the lens light. If a detected source is spatially and spectrally extended, this is a strong indication of an emission line, which is then processed for identification.

The candidate emitters spectrum is extracted within a 2\,arcsec diameter aperture, and at $\pm$\,125\,\AA\ around the brightest detected pixel. This range is chosen to contain [{\sc O\,iii}]\,4959\,\AA\ if the lead detection is [{\sc O\,iii}]\,5007\,\AA, whilst avoiding H\,$\beta$ to simplify the emission line fitting. (The range will also contain [N\,{\sc ii}] for H\,$\alpha$.) The extracted spectrum is then fit with a single, double and triple gaussian, with appropriate peak ratios and separation for [{\sc O\,ii}], [{\sc O\,iii}], and H\,$\alpha$+[N\,{\sc ii}]. We perform a chi-square minimisation to select the best fit identification, and measure a redshift. 

After these detections have been carried out for each 40\,\AA\ segment of the datacube, we matched detections spatially in order to combine sources with multiple emission lines at a consistent redshift, and included a step to associate the single gaussian lines to other identified lines (generally this matches H\,$\beta$ to detected [{\sc O\,iii}]). We do not specifically fit the asymmetric Lyman--$\alpha$ profile, as these will be best fit with a single gaussian with a wide FWHM, and so will be included in the sample without more complex modelling.

\begin{figure*}
	\centering
	\includegraphics[width=0.9\linewidth]{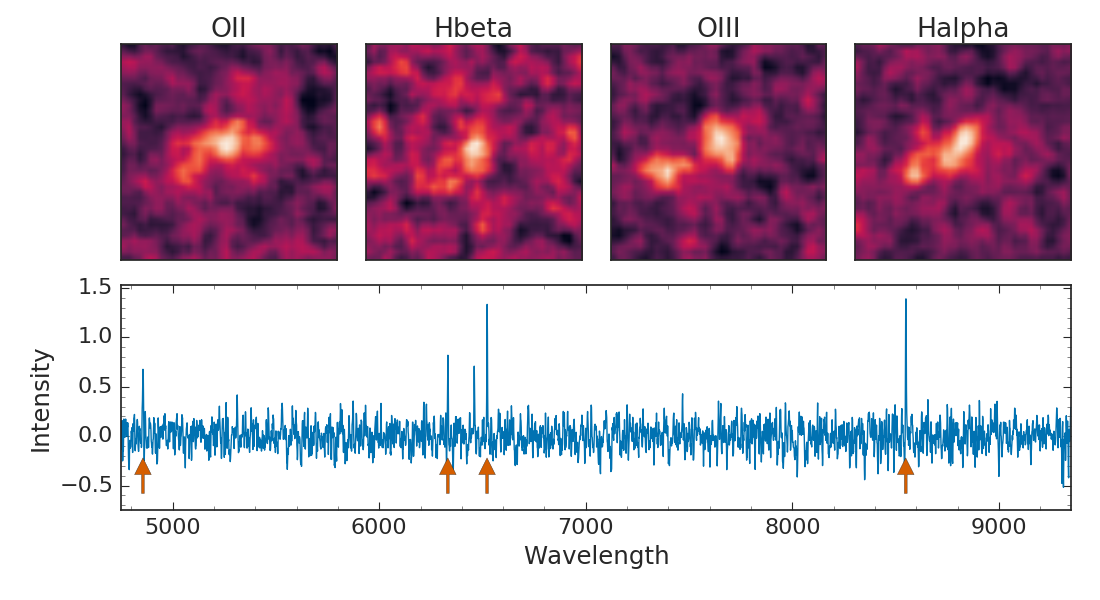}
	\caption{An example emitter from a MNELLS observation. The upper panels show the spatial extent of the emission in narrow-band slices around [{\sc O\,ii}], H\,$\beta$, [{\sc O\,iii}] and H\,$\alpha$, which tend to be the strongest optical emission lines. The panel size is 8\,$\times$\,8\,arcsec. Below is the spectrum extracted within a 2\,arcsec diameter aperture, at redshift 0.3024. The emission lines extracted for the narrowband images are indicated with orange arrows.}
	\label{fig:exoutput}
\end{figure*}

Finally the candidate lines are manually inspected and either verified as sources or rejected. Sources within a 10\,arcsec radius of the candidate lens centre are recorded separately. This minimises the time per datacube on a first pass for lensed images. In Figure \ref{fig:exoutput}, we show an example of the outputs which were visually inspected. Narrowband images of the respective positions for the four strongest optical emission lines ([{\sc O\,ii}], H\,$\beta$, [{\sc O\,iii}], and H\,$\alpha$) are displayed in the top panels. Inspecting the panels will show those detections which appear false in the spectral domain due to their lack of a clear peak, or spatial extent due to a residual (see Appendix \ref{app:rejem} for an example of an excluded detection).

The end results of the processing described in this section, is a final catalogue of visually screened emission line sources for each datacube.

\begin{table*}
	
	\caption{The results of the MNELLS survey. We include the expsoure time, galaxy redshift and absolute K$_s$ magnitude. We state for each galaxy if there is a multiple-- or single-- imaged lensing system. Where either of these are possible, we report the distance to any emitters within 10\,arcsec. We use `large separation' to describe systems with an emitter more distant than 6\,arcsec ($\sim$\,2\,\rein), but within 10\,arcsec. Single-imaged denotes a galaxy with an emitter within 6\,arcsec. J13522521-3456009, and J15105568-1128477 were only observed for a single exposure.}
	
	\label{tab:MNELLS}
	\begin{tabular}{|l|c|c|c|c|c|c|}
			\hline Galaxy (2MASX) &  t$_{\rm exp}$ & $z$ & K$_{\rm iso}^{\rm 2MASS}$ & Total Emitters & Single-Imaged & Ang. Sep. (arcsec) \\ 
			\hline 
			J00031127-5444588 & 3060 & 0.03258 & --25.456 & 22 & large sep. & 9.7   \\ 	
			J00585131-1628092 & 3060 & 0.05408 & --26.484 & 15  & -- & --   \\ 
			J01455353-0656086 & 3060 & 0.05229 & --25.460 & 14  & large sep. & 7.6, 7.7  \\ 	 
			J02023082-5055539 & 3060 & 0.02148 & --25.127 & 9  & Yes & 1.9,3.4,4.7,5.6,7.6,9.3  \\
			J03023835-6032198 & 3060 & 0.05425 & --25.494 & 6  & -- & --   \\	 	
			J05480837-4724177 & 3060 & 0.05166 & --25.939 & 6  & -- & --   \\ 			 	 
			J11530823-3233574 & 3060 & 0.02712 & --25.331 & 11  & large sep. & 9.7   \\ 	 
			J12332514-3121462 & 3060 & 0.05194 & --26.395 & 19  & Yes &  5.1, 7.0, 7.1, 8.7 \\ 
			J13522521-3456009 & 1530 & 0.03824 & --25.497 & 25 & Yes & 5.0,7.0 \\ 
			J15105568-1128477 & 1530 & 0.02495 & --26.011 & 9  & -- & --   \\ 		 
			J18322937-6017262 & 3060 & 0.05170 & --25.564 & 15  & large sep. & 7.2   \\
			J19163258-4012332 & 3060 & 0.01858 & --25.676 & 7 & large sep. &  9.1  \\ 		 
			J19281700-2931442 & 3060 & 0.02432 & --25.456 & 15 & large sep. &  6.6, 7.7, 9.3  \\ 
			J21075218-4710445 & 3060 & 0.01620 & --24.969 & 5 & large sep. &  9.7  \\
			J21293744-2111443 & 3060 & 0.03563 & --26.116 & 13 & -- & --   \\ 
			J23184637-1023575 & 3060 & 0.03170 & --25.347 & 7 & large sep. & 6.6   \\ 		
			\hline 
	\end{tabular} 
\end{table*}

\begin{table*}
	\caption{The results of our lens search within the MUSE archive. We define each galaxy by its 2MASS ID, and in the second column the target object from the MUSE data header. For each galaxy we report the number of emitters, the exposure time, and if there is evidence for a multiple-- or single-- imaged lens system. The distance to the nearest emission line source within 10\,arcsec is noted, or the distances to the multiple images. We also note two `potential' systems which have single emission line detections which we do not find compelling.}
	\label{tab:archive}
	\begin{tabular}{|l|c|c|c|c|c|c|c|}
		\hline 2MASX & Target (header) & t$_{\rm \exp}$ & Tot. Emitters & Multiply-Im. & Single-Im & Ang. Sep. (arcsec) & Programme ID \\
		\hline  
		
		J00561610-0115197 & PGC003342 & 15660 & 39 & -- & large sep. & 8.4  & 095.B-0127(A) \\ 
		J01024177-2152557 & LSQ13cmt & 2805 & 5 & -- & -- & --  & 099.D-0022(A) \\ 
		J01145760+0025510 & PGC004500 & 5180 & 12 & -- & Yes & 4.7  &  094.B-0592(A),099.B-0242(A) \\ 
		J01260057-0120424 & 3C040 & 1200 & 11 & -- & Yes & 5.4 & 099.B-0137(A) \\ 
		J02001493+3125457 & NGC777 & 2700  & 20 & -- & -- & -- & 097.A-0366(B) \\ 
		J02021730-0107405 & PGC007748 & 6390 & 20 & -- & Yes & 3.5, 4.2  &  094.B-0592(A) \\ 
		J02150461-2046037 & SN2006ot & 2220 & 8 & -- & -- & -- &  095.D-0091(B) \\ 
		J02431504+3225300 & NGC1060 & 2700  & 15 & -- & -- &  -- & 097.A-0366(B) \\ 
		J04333784-1315430 & A0496 & 9000 & 38 & -- & -- & --  & 094.B-0592(A),095.B-0127(A) \\ 
		J04035024-0239275 & LSQ13cwp & 2805 &  22 & Yes & -- & 1.17, 1.71 & 098.D-0115(A) \\ 
		J05005065-3839169 & LSQ12fvl & 2805 & 1 & -- & -- & --  &  099.D-0022(A) \\ 
		J05571255-3728364 & AS555 & 2700 & 17 & Potential & -- & -- &  094.A-0859(A) \\ 
		J06004111-4002398 & A3376 & 4680  & 20 & -- & large sep. & 8.4  &  094.B-0592(A)  \\ 
		J06273625-5426577 & A3395 & 16590  & 29 & -- & Yes & 4.4 &   094.B-0592(A),096.B-0062(A) \\ 
		J12542222-2900466 & PGC043900 & 6840 &  10 & -- & -- & -- &  097.B-0776(A) \\ 
		J12571157-1724344 & A1644 & 2700 & 7 & -- & -- & -- &  094.A-0859(A)  \\ 
		J13232900-3150392 & PGC046785 & 8280 & 33 & -- & large sep. & 6.8  & 096.B-0062(A),097.B-0776(A) \\ 
		J13240669-3140118 & PGC046832 & 3240 & 25 & -- & -- & -- &  096.B-0062(A),097.B-0776(A) \\ 
		J13242275-3142239 & PGC046860  & 2880 &  11 & -- & Potential & 2.7 &  098.B-0240(A) \\ 
		J13272961-3123237 & PGC047154  & 3240 & 18 & -- & -- & -- &  095.B-0127(A),096.B-0062(A) \\ 
		J13275493-3132187 & PGC047197 & 2880 & 12 & -- & -- & -- &  095.B-0127(A),096.B-0062(A) \\ 
		J13275688-3129437 & PGC047202 & 1440 & 16 & -- & -- & --  &  095.B-0127(A) \\ 
		J13283871-3120487 & PGC047273 & 5040 & 15 & -- & large sep. & 7.1 &  097.B-0776(A) \\ 
		J13292810-3133048 & PGC047355 & 3240 & 19 & -- & large sep. & 8.9  & 096.B-0062(A),097.B-0776(A) \\ 
		J13303277-3134339 & PGC047467 & 2880  & 19 & -- & -- &  -- &  098.B-0240(A) \\ 
		J13312752-3149140 & PGC099188 & 8640  & 27 & -- & -- & -- &  098.B-0240(A),099.B-0148(A) \\ 
		J13320334-3146430 & PGC047590 & 5940 & 43 & -- & Yes & 4.7 & 098.B-0240(A),099.B-0148(A) \\ 
		J13333473-3140201 & PGC047752 & 3960 & 19 & -- & large sep. & 7.7  &  096.B-0062(A) \\ 
		J13472838-3251540 & PGC048896 & 13920 & 20 & -- & -- & -- &  097.B-0776(A) \\ 
		J14014186-1136251 & PGC049940 & 2160 &  7 & -- & -- & --  &  097.B-0776(A) \\ 
		J14543146+1838325 & A1991 & 2700 & 8 & -- & -- & -- &  094.A-0859(A) \\ 
		J15164448+0701180 & A2052 & 7485 & 38 & Yes & -- & 7.4, 13.2  &  097.B-0766(A) \\ 
		J20515691-5237473 & PGC065588 & 14310 & 41 & -- & -- & -- &  097.B-0776(A),099.B-0148(A) \\ 
		J23135863-4243393 & AS1101  & 2700 & 16 & -- & Yes  & 3.6 &  094.A-0859(A)  \\ 
		J23363057+2108498 & A2626 & 1800 & 7 & -- & Potential & 3.5 &  095.A-0159(A) \\ 
		J23570068-3445331 & A4059 & 10080 & 33 & Yes & -- & 8.6, 9.8  & 094.B-0592(A) \\ 

		\hline 
	\end{tabular} 
\end{table*}

\section{Identified/Candidate Lenses}
\label{sec:ICL}

In this Section we will detail the results of the line emission search on a total of 52 galaxies. The results are summarised in Table \ref{tab:MNELLS} for MNELLS, and Table \ref{tab:archive} for the archival search. 

We use three main criteria for any galaxy we label as a multiply-imaged lens. The first is a secure redshift for each image, from either multiple consistent emission lines or a clear [{\sc O\,ii}] doublet. The second is a configuration which has a strong resemblance to theoretical lens systems modelled with a singular isothermal sphere (SIS) or singular isothermal ellipsoid (SIE) parametric model. The third is any velocity offset between each image must be small ($\nu_{\rm offset} <$\,100\,km\,s$^{-1}$), and similar emission line ratios. The presence of weaker, less commonly observed emission lines, (i.e O{\sc i}, H$\gamma$, H$\delta$, [Ne\,{\sc iii}] or He\,{\sc i}) in both spectra are also strong indicators of a common source. For a close-projected single-imaged system, only the first is relevant.


In the combined sample, four galaxies show evidence for double-imaged sources, with confirmation from multiple emission lines. These are 2MASXJ04035024-0239275 (here-after J0403-0239) an isolated elliptical, Section \ref{sec:J0403}, and three BCGs, from the clusters A4059, A2052 and AS555 i.e. 2MASXJ23570068-3445331, 2MASXJ15164448+0701180, 2MASXJ05571255-3728364 respectively. Detailed discussion and analysis of these `cluster-lenses' is presented in Section \ref{sec:MIC}. 

During our systematic search, we recorded separately any emitters discovered within 10\,arcsec of the galaxy centres. For a given lens, the region on the image-plane within which a source can form multiple images is 2\,$\times$\,\rein. For our sample, the typical \rein\ is 2--3\,arcsec (for an isothermal sphere model with velocity dispersion 260--320\,km\,s$^{-1}$). We discovered nine galaxies with close-projected but apparently single-imaged sources within 6\,arcsec. These galaxies are analysed to constrain the maximum lensing mass which produces no detectable counter-image, and hence the lens galaxy IMF \citep*[`Upper Limit Lensing'][]{Smith2018}. We describe these close-projected systems in Section \ref{sec:SIC}.


Whilst searching for lenses in datacubes with such a large FoV, there are many cases of multiple clustered background emitters at the same redshift. In Section \ref{sec:MEG}, we explain the criteria we use to exclude such systems.

\subsection{Double-image lens J0403-0239}
\label{sec:J0403}

During the archival lens search, we discovered the strong-lens system J0403-0239, which is a massive ($\sigma$\,=\,314\,km\,s$^{-1}$) ETG without any nearby neighbours, lying at $z$\,=\,0.06655. The pair of background emitters lie at $z$\,=\,0.195, and the system has an \rein\ of 1.47\,arcsec. This \rein\ probes the stellar dominated galaxy core, and in \citet{Collier2018b} we reported the implications of this system on the IMF within massive ETGs. We found this galaxy to have a lightweight IMF, under the assumption of a typical old stellar population. The discovery of this lens galaxy was also independently reported by \citet{Galbany2018}, who describe the potential presence of a second younger stellar population which may affect the conclusion of a lightweight IMF. 

\begin{figure*}
	\centering
	\includegraphics[width=\linewidth]{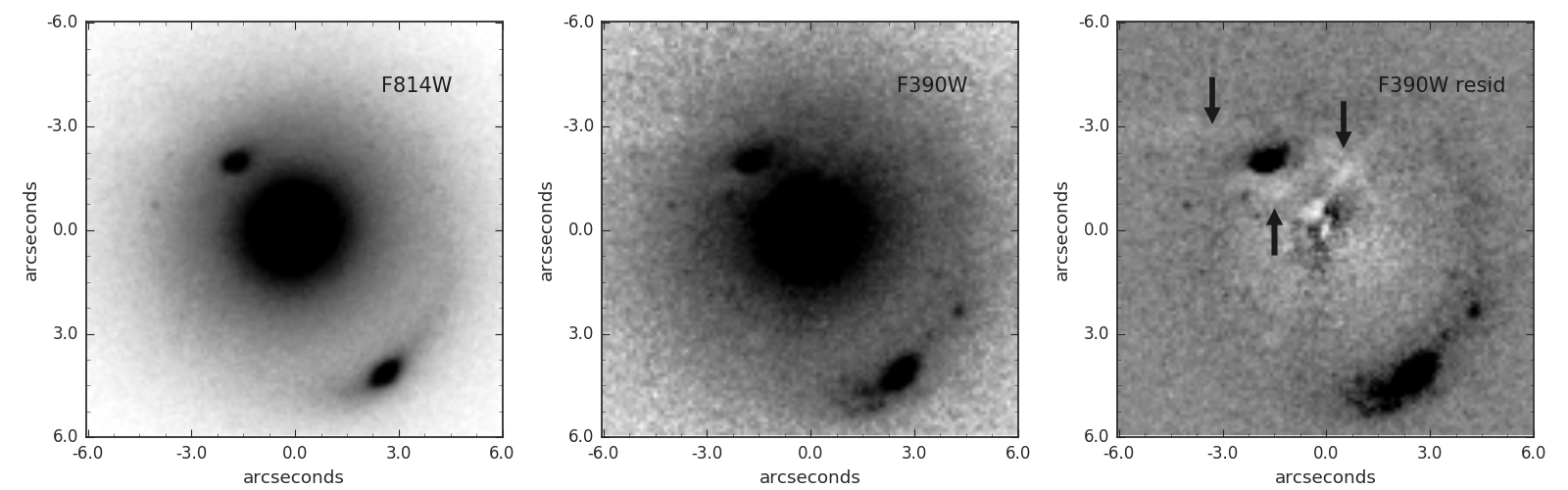}
	\caption{The recently acquired HST data for J0403-3055. {\it Left}: F814W HST data, showing a smooth lens light profile. The lensed images appear smooth, and a hint of a faint arc can be seen about the outer image. {\it Middle}: F390W HST data, displaying structure in the outer image. {\it Right}: A lens light subtracted image. The strongest evidence for a recent interaction is the clumpy residual which can be traced inside to the inner image, and a small plume more distant than the inner image (marked by the arrows).}
	\label{fig:J0403hst}
\end{figure*}

We have since acquired HST observations (PI: Smith) of J0403-0239, using Wide Field Camera 3 (WFC3/UVIS). We observed with the F814W and F390W filters, for 1040, and 3900\,s respectively. The imaging is shown in Figure \ref{fig:J0403hst}. The lens has a smooth light distribution in the red (F814W), and extended lensed images with well-defined clumpy structure at shorter wavelengths (F390W). However in the F390W imaging, there are signs of irregular patchy dust obscuration running close to the lens galaxy near to the inner image (see right panel, Figure \ref{fig:J0403hst}). This adds to evidence from the MUSE datacube and wide-field imaging that suggest this galaxy has been recently perturbed.


We perform lens analysis with {\sc pyautolens} \citep{Nightingale2018} using the F390W HST images to exploit the lensed image structure. Modelling the source with either a S{\'e}rsic model, or a full pixelised inversion, we find a compact source as a best fit solution to the lens plane image positions and shape. Therefore the only change in constraints relative to the earlier analysis is that the peak pixel \rein\ is 1.49\,arcsec, which is 0.5 per cent larger than the same measurement in the MUSE data. This changes our mass-measurement by less than 2 per cent ($\alpha$\,=\,1.17\,$\pm$\,0.17). Hence our previous conclusions are unchanged. 

In addition to the new HST imaging, we observed J0403-0239 with ESO/VLT FORS2. The wavelength coverage of the new data (3660\,--\,5110\,\AA) is blueward of MUSE (4750\,--\,9350\,\AA), and hence contains the higher order Balmer series, and the Ca\,{\sc ii} doublet. These absorption features are key to precisely constraining the age of a stellar population. With the FORS2 data we can test our assumption that J0403-5055 has a single, old stellar population against the old population `frosted' with a considerably more recent starburst which is measured from the MUSE data with STARLIGHT \citep{Galbany2018}.

\begin{figure*}
	\centering
	\includegraphics[width=\linewidth,trim = 0 1.0cm 0 0]{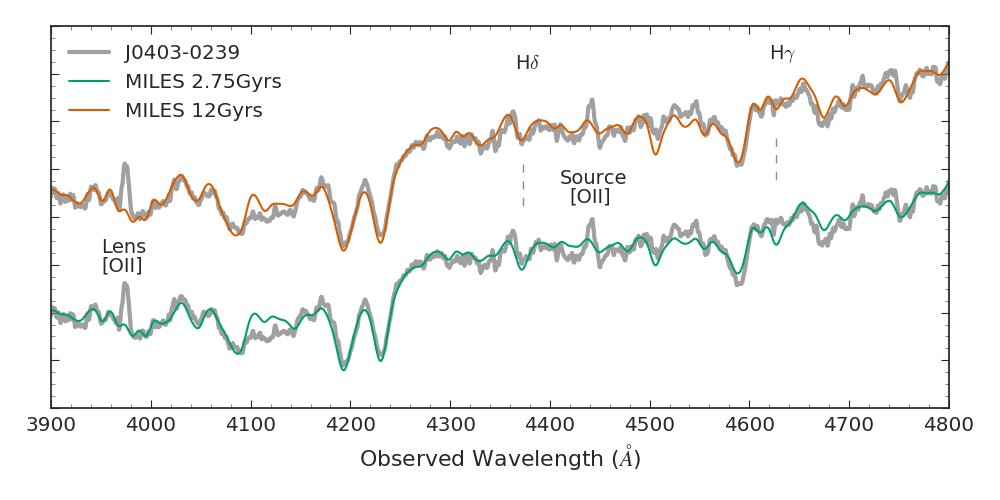}
	\caption{The recently acquired ESO/VLT FORS2 data for J0403-3055 extracted within \rein, is displayed in grey. The data are not flux calibrated, and hence we match the continuum shape to an old stellar population. We compare the data with two differently aged SSPs from the MILES library \citep{Vazdekis2010}. The upper, is an old, 12Gyr population which is $\alpha$-enhanced and metal-rich. The lower, is a 2.75Gyr population as suggested by \citet{Galbany2018}. We label the H\,$\delta$ and H\,$\gamma$ absorption features, which appear better matched by the older population.}
	\label{fig:J0403FORS}
\end{figure*}

We extract a spectrum within an aperture matched to the \rein, covering the wavelength range 3900\,--\,4800\,\AA\ to include the high-order Balmer series, see Figure \ref{fig:J0403FORS}. As a simple comparison, we overlay the FORS2 spectrum with two different age SSPs from the MILES models \citep{Vazdekis2010}, assuming a bimodal IMF with a slope of -1.30. We select models which are $\alpha$-enhanced, and metal-rich with ages of 2.75\,Gyrs \citep[][measured a 2.6\,Gyr luminosity-weighted population]{Galbany2018}, and 12\,Gyrs \citep[an old population was assumed by][]{Collier2018b} which are redshifted to 0.066, and smoothed to $\sigma$\,=\,314\,kms$^{-1}$. 

The FORS2 spectrum does not exhibit the strong H$\gamma$ and H$\delta$ absorption expected from a $<$\,3\,Gyr stellar population, nor the strong high-order Balmer absorption characteristic of composite populations with $<$\,1\,Gyr components. Despite the evidence for a recent interaction or accretion event in J0403-5055, the FORS2 spectrum supports the assumption in \citet{Collier2018b} of an old stellar population inside the \rein. A more rigorous analysis will be presented elsewhere.


\subsection{Multiply-imaged cluster lenses}
\label{sec:MIC}

In addition to J0403-0239, we discover multiply-imaged sources behind two cluster BCGs (2MASXJ23570068-3445331 and 2MASXJ15164448+0701180). These detections are shown in Figures \ref{fig:A4059} and \ref{fig:A2052}. We also report a third potential lens, 2MASXJ05571255-3728364, in Figure \ref{fig:S555}. These are each among the lowest redshift clusters known to have multiply-imaged sources.

\subsubsection{2MASX J23570068-3445331}
\label{sec:A4059}

The first candidate lens, J2357-3445, lies at $z$\,=\,0.0491, and is the cluster BCG of A4059. This cluster lies at $z$\,=\,0.0487, with a size of $R_{500}$\,$\simeq$\,0.96\,Mpc and a mass, $M_{500}$\,$\simeq$\,2.67\,$\times$\,10$^{14}$\,M$_{\odot}$ taken from the `meta-catalogue of X-ray detected clusters of galaxies' \citep[MCXC, ][]{Piffaretti2011}. This cluster has been subject to multi-waveband observations, which could provide additional information to describe this lensing system. 
In HST/WFPC2 F814W imaging, \citep{Choi2004}, the presence of a dust lane is clear. Furthermore, large plumes of filamentary nebular emission at the cluster redshift \citep[][]{McDonald2010} is present in the MUSE data.

\begin{figure*}
	\centering
	\includegraphics[width=\linewidth, trim = 0 1cm 0 0 ]{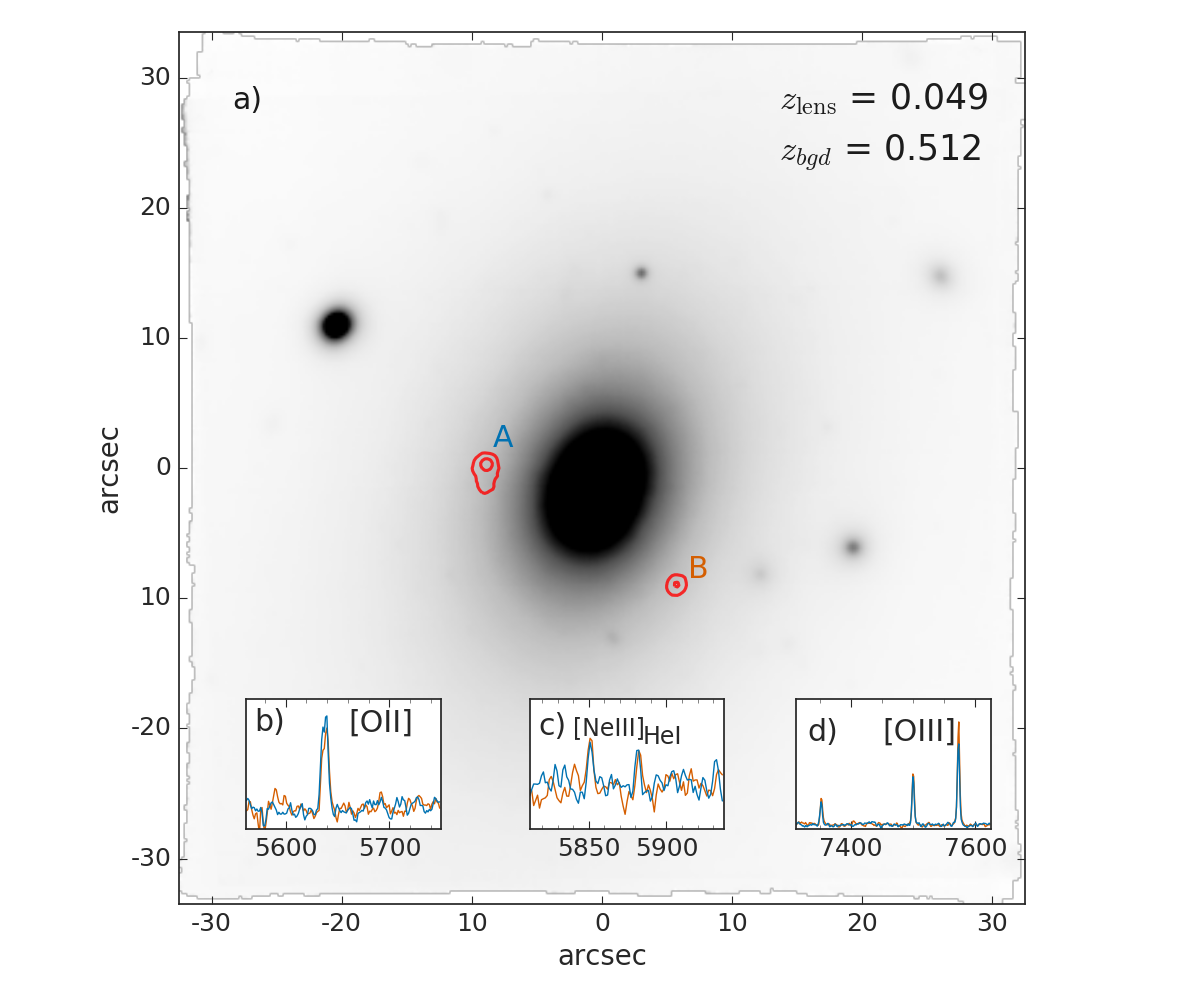}
	\caption{The MUSE data for the BCG J2357-3445, and the extracted spectra of the multiply-imaged background emitter. Panel a) The MUSE data, collapsed over the full wavelength range, for J2357-3445. Contours at the peak [{\sc O\,iii}] emission for the background emitters are displayed in red. Panel b), c) and d) show the emission from the two images with their negligible velocity offset in [{\sc O\,ii}], [Ne\,{\sc iii}], and [{\sc O\,iii}] respectively, which confirm the single source of origin.}
	\label{fig:A4059}
\end{figure*}

We detect two background emitters in the MUSE datacube separated by 17.15\,arcsec which are at $z$\,=\,0.512 (see Figure \ref{fig:A4059}a). In the HST data, there is a very faint source coincident with image A, though it cannot be unambiguously determined as related. Though the separation is large, the spectra are similar, with consistent line ratios of [{\sc O\,iii}], H\,$\beta$ and [{\sc O\,ii}], in Figure \ref{fig:A4059}b,c,d. A strong suggestion of these sharing a common source, and not being two different background galaxies, is the presence of weak [Ne\,{\sc iii}] $\lambda$3869 and He\,I $\lambda$3888 lines, which are present in both spectra with similar line ratios. Neither of these lines is commonly observed in galaxy spectra, hence these images are likely to originate from a single background source. Image A appears slightly extended, which cannot be ruled out to be present in image B. Due to the small velocity offsets, and rare emission lines, this system is labelled a lens.


The very large angular separation means J2357-3445 is a weaker tool for investigating the IMF, as additional information is required to disentangle the dark and stellar matter once the lensing mass is estimated. Indeed, in such an environment, it is expected the cluster DM will dominate the lensing mass.



Fitting an SIE profile to images A and B, measures a lensing mass within the half image-separation of 1.79\,$\times$\,10$^{12}$\,M$_{\odot}$. This is over six times larger than expected from the stellar mass alone, for a MW-like IMF, which indicates that DM is likely dominating the lensing mass. A more detailed discussion is presented in Section \ref{sec:LAMIS}.



\subsubsection{2MASXJ15164448+0701180}
\label{sec:A2052}

J1516+0701 is a massive elliptical galaxy, at $z$\,$=$\,0.0345. It is the BCG of A2052 ($z$\,=\,0.0355), which has an X-ray measured $M_{500}$, and $R_{500}$, of 2.5\,$\times$\,10$^{14}$\,M$_{\odot}$ and 0.95 Mpc \citep{Piffaretti2011}. There is some extended emission in the datacube, indicating that this is an active galaxy, and complicating the detection of background objects.

\begin{figure*}
	\centering
	\includegraphics[width=\linewidth, trim = 0 1cm 0 0 ]{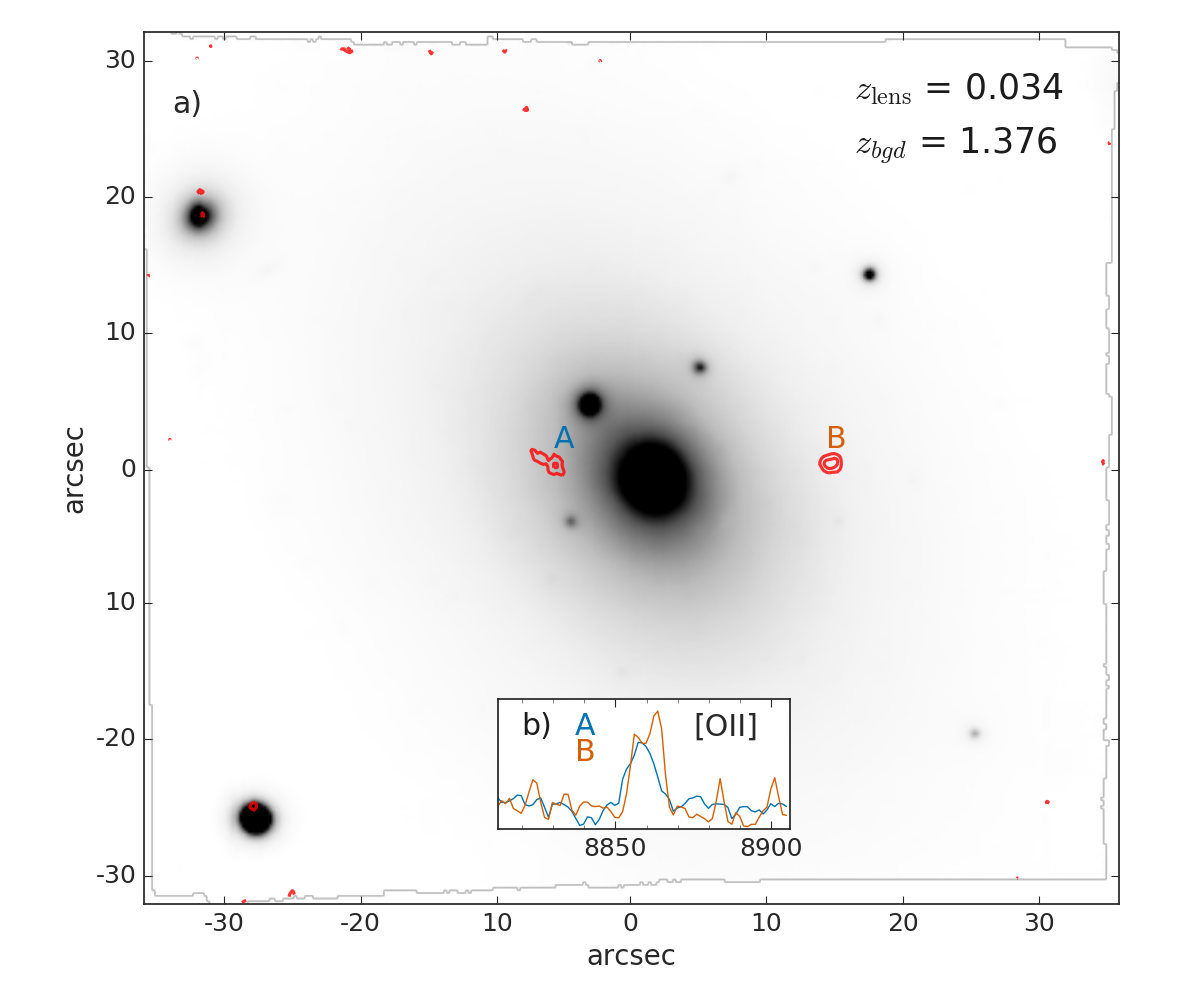}
	\caption{The MUSE data for the BCG J1516+0701, and the extracted background emitter spectra. Panel a) The MUSE data for J1516+0701 collapsed over the full wavelength range, with red contours denoting the position of the $z$\,$=$\,1.377 background emitters. Panel b) The [{\sc O\,ii}] emission from images A and B overlaid. There is very little velocity offset, however A only hints at a complete doublet structure. The structure in image A cannot be ruled out in image B.}
	\label{fig:A2052}
\end{figure*}

The pair of emitters lie west-east in Figure \ref{fig:A2052}a), and are separated by 19.4\,arcsec. The closest image, A, is located 7.4\,arcsec west, and B is 13.2\,arcsec east from J1516+0701, and they are separated by $<$\,50\,km\,s$^{-1}$ in velocity space. The emitters were detected from the [{\sc O\,ii}] doublet at $z$\,=\,1.376, see Figure \ref{fig:A2052}b. There are no other spectral lines in the MUSE wavelength coverage. Both images appear to share the same structure, with two distinct clumps in A, and potentially the same in B. There are archival HST observations (WFPC2 F814W, 6500s; PI:Geisler) however we cannot detect any obvious counterparts to the MUSE detections. Due to the similar extended structure and clear doublet in image B, we label this a strong-lensing system.

This system is similar to J2357-3445, being located in a cluster environment, however the extended emission maybe indicates the presence of star formation or an active galactic nuclei. This makes the system even more complex for investigating the IMF as selecting a reference mass-to-light ratio (\MLref) is more uncertain. 


From SIE paramaterised lens-modelling, the mass within half the image separation is 1.65\,$\times$\,10$^{12}$\,M$_{\odot}$. This is seven times larger than expected from a stellar population alone from a MW-like IMF. Therefore the lensing mass for the galaxy probably has a significant contribution from DM. A more detailed discussion is presented in Section \ref{sec:LAMIS}.


\subsubsection{2MASXJ05571255-3728364}
\label{S555}

The final and least secure cluster candidate is J0557-3728, the BCG of AS555, which lies at redshift 0.0448. The cluster has a redshift, $z$\,=\,0.0440, with an X-ray measured $M_{500}$ and $R_{500}$ of 0.97\,$\times$\,10$^{14}$\,M$_{\odot}$ and 0.69\,Mpc \citep{Piffaretti2011}. There is a significantly smaller companion galaxy separated by $\sim$\,3.2\,arcsec, but no similarly sized nearby galaxies within 2 arcmin. The emitters were discovered via strong [{\sc O\,ii}] emission, at redshift 0.87. However, the candidate lensed images do not follow a classic lensing configuration, A is located 20.7\,arcsec south-south-east, with C only 3\,arcsec further, and B is located 16.2\,arcsec north, meaning that the images do not intersect the lens galaxy. There is a small velocity offset of 80\,km\,s$^{-1}$ between A, and C, and between B and C. We tentatively label this system as a lens.

\begin{figure*}
	\centering
	\includegraphics[width=\linewidth, trim = 0 1cm 0 0 ]{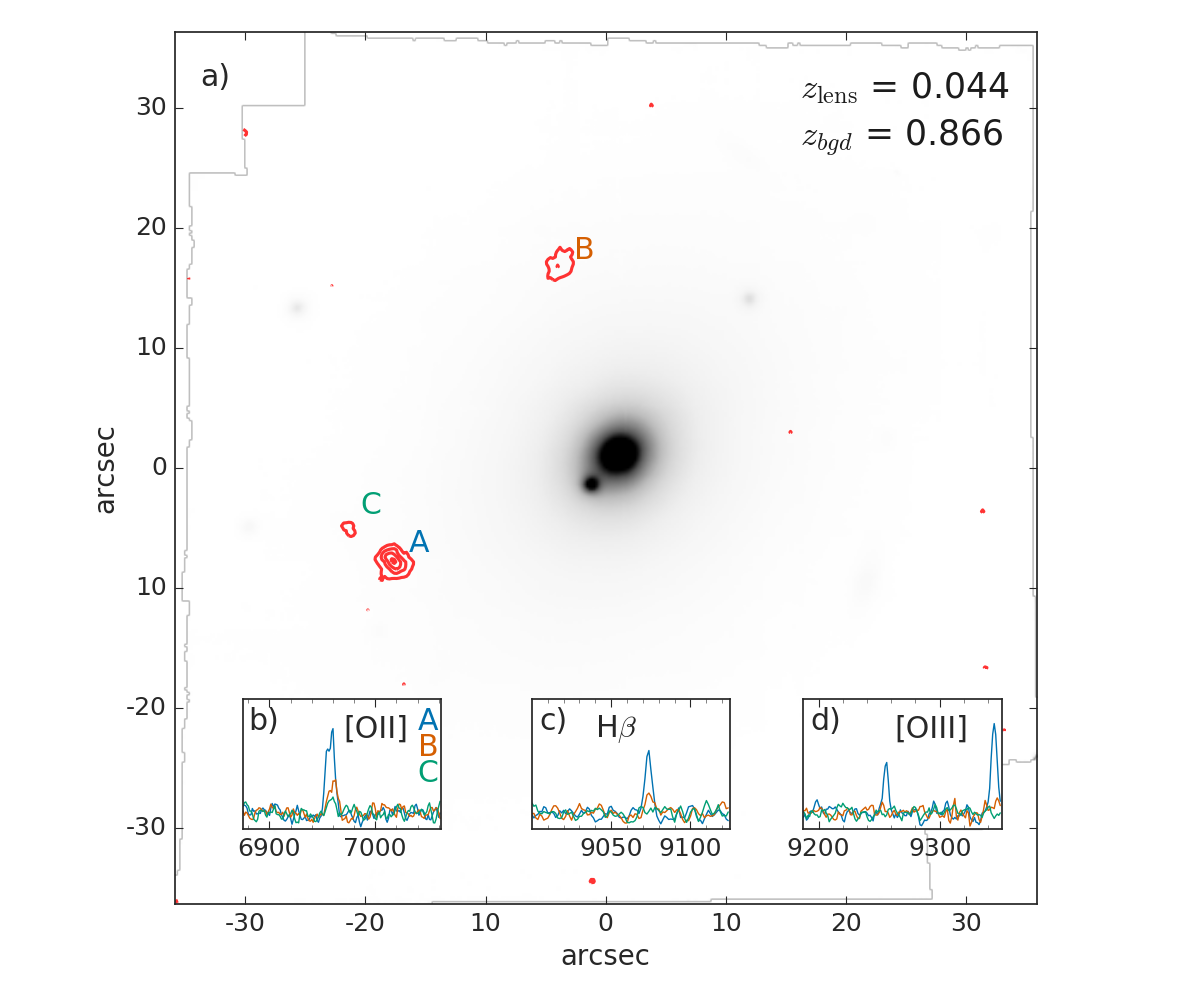}
	\caption{The MUSE data for the BCG J0557-3728, and the extracted background emitter spectra. Panel a) The MUSE data for J0557-3728 collapsed over the full wavelength range which shows the close, small companion galaxy. Overlaid on top are the contours of the [{\sc O\,ii}] emission for the candidate lensed images. Panels b), c) and d) show the strongest emission lines from the three clumps labelled in a).}
	\label{fig:S555}
\end{figure*}

This system is significantly more complex than the previous two cluster lenses, and the exposure time is relatively short (2700 seconds) which limits the detection of faint sources. The surrounding environment does not appears to suggest that a large external shear is the cause of the non-standard configuration. One potential explanation could be that J0557-3728 is offset from the centre of the AS555 cluster DM. An alternative explanation originates from the line strength ratios between the different images, Figure \ref{fig:S555}b,c,d. Common lines (i.e. H\,$\beta$ and [{\sc O\,iii}]) are stronger, and weaker in varying ratios between the three images. This may be due to differential magnification of components of a single source. This may be explained by the source is crossing a caustic line, and the existence of image C potentially adds evidence for this explanation, as it may be part of image A. Due to the complex nature of this system, and few constraints, we do not attempt to model this system for the purpose of constraining the IMF.


Further data, such as deeper MUSE observations could help to confirm the lensing configuration, ruling out or discovering additional faint counter-images which could be used to constrain this system.


\subsection{Multiple close emitters}
\label{sec:MEG}

Within any search for multiply-imaged lensing systems, there will be cases for which the observations do not provide conclusive evidence. Often the distinguishing features between a lensing and non-lensing interpretation require significant case-by-case analysis. As summarised in Section \ref{sec:ICL} the three main criteria are, the redshift quality, the lensing configuration, and the velocity offset between images. To illustrate the decision process we show two examples of rejected systems in Appendix \ref{app:MCE} along with an explanation of the evidence which led to them being rejected. The first case is rejected due to a significant velocity offset, and varying lines ratios. The second is excluded as there is no apparent counter-image, and the sources are located outside of the expected multiple-imaging regime.

\subsection{Singly-imaged candidates}
\label{sec:SIC}

\begin{figure*}
	\centering
	\includegraphics[width=1.0\linewidth]{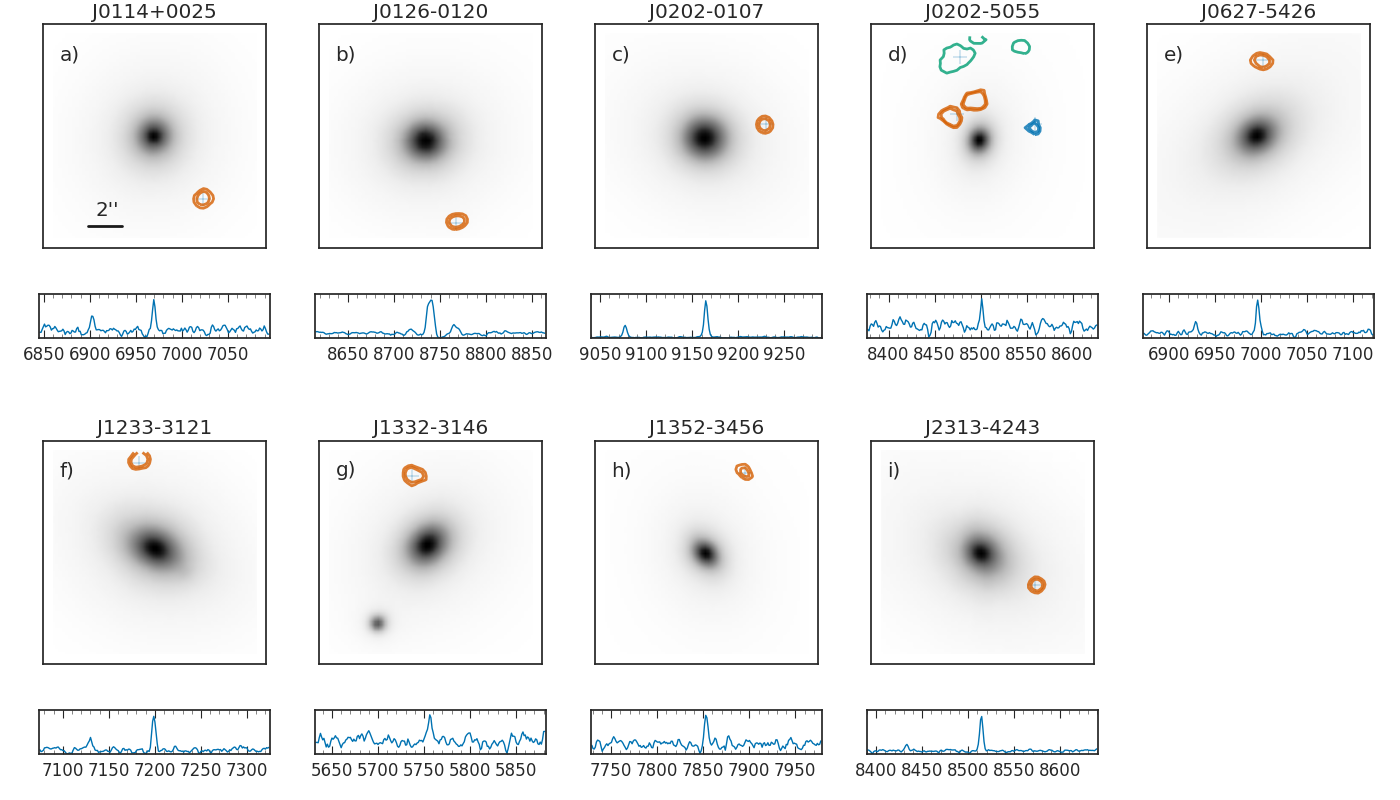}
	\caption{The single-imaged systems as described in Section \ref{sec:SIC}. Each frame consists of the collapsed MUSE datacube across the full spectral range, overlaid with a orange contour of the brightest peak of emission for the single-imaged closely projected background emitter. All panels have the same scale, and are 12"\,$\times$\,12". Underneath each MUSE frame is the spectrum extracted at peak of the emission, and only with the brightest peak/doublet shown. In the majority of cases this is the [{\sc O\,iii}] doublet. In the case of J0202-5055 we show all three emitters, the smallest angular separation in orange, then blue, and then green (The green contours have multiple regions, as there is a group of clustered emitters at the same redshift).}
	\label{fig:sl}
\end{figure*}

Although multiply-image systems provide the strongest constraints, there is also information in estimating upper-limits for \ML\ of a lensing system in which only a single closely-projected emitter is detected \citep[e.g.][]{Shu2015,Smith2018}. Accounting consistently for such systems can help mitigate lensing selection bias. Here, we follow the \citet{Smith2018} methodology, but with a larger and more uniform sample. We select singly-imaged, but closely projected `lenses' as those galaxies with an emitter within 6\,arcsec of the galaxy centre, shown in Table \ref{tab:MNELLS} and \ref{tab:archive}. Within our dataset we discovered nine such candidates. Each detection is visually inspected in the spatial and spectral domain to confirm the emission lines. Each of the nine (six and three from the archival and targeted searches respectively) candidates is displayed in Figure \ref{fig:sl}, and are summarised in Table \ref{tab:SL_can}. These are the best candidates for single lensing analysis. We exclude 2MASXJ23363057+2108498, and 2MASXJ13242275-3142239 which each have sources within 6\,arcsec in Table \ref{tab:archive}. In each of these cases the source only has a single line detected which we do not find compelling (due to the lack of clear [{\sc O\,ii}], or Ly$\alpha$ structure), and therefore do not have a confidently identified redshift. 

In this subsection we outline each candidate, and then in Section \ref{sec:LASIS} show the upper-limit lensing analysis results and make comment on whether any of these systems are promising candidates for follow-up observations.

\begin{table*}
	
	\caption{The single-image candidates. For each galaxy we list the galaxy redshift, the redshift of the background source, and the separation in\,arcsec. The position angle is denoted in degrees north through east.}
	
	\label{tab:SL_can}
	\begin{tabular}{|c|c|c|c|c|}
			\hline Galaxy (2MASX) & $z_{\rm gal}$  & $z_{\rm bgd}$ & Separation (arcsec) & Position Angle (degrees) \\ 
			\hline 			 	 
			J01145760+0025510 & 0.04490 & 0.390 & 4.70 & --141.8   \\ 
			J01260057-0120424 & 0.01824 & 0.332 & 5.39 & --164.9 \\
			J02021730-0107405 & 0.04276 & 0.830 & 4.20 & --103.0  \\		 
			J02023082-5055539 & 0.02148 & 0.295 (1.29, 0.907) &  1.92 (3.37, 4.68) & 50.0 (--78.0,--27.3) \\ 
			J06273625-5426577 & 0.04856 & 0.3971 & 4.40 & --2.5  \\ 
			J12332514-3121462 & 0.05194 & 0.4374 & 5.14 & 11.7  \\  
			J13320334-3146430 & 0.04372 & 0.890 & 4.24 & --102.7 \\	
			J13522521-3456009 & 0.03824 & 0.1962 & 5.06 & --23.5 \\ 
			J23135863-4243393 & 0.05640 & 0.700 & 3.60 & --119.2  \\ 	 	
			\hline 
	\end{tabular} 

\end{table*}

\subsubsection{2MASXJ01145760+0025510}
\label{A0168}

The BCG of A0168, J0114+0025 has $z$\,=\,0.04490 ($\Delta v_{\rm cluster}$\,=\,70\,kms$^{-1}$). The observation is an archival MUSE-DEEP datacube with a PSF of 1\,arcsec. MCXC X-ray data \citep{Piffaretti2011}, tabulates the cluster $R_{500}$ at 0.75\,Mpc, and $M_{500}$\,=\,1.25\,$\times$\,10$^{14}$\,M$_{\odot}$. The background emitter is separated by 4.7\,arcsec to the south-west, and so is likely at the outskirts of the strong-lensing regime. The collapsed MUSE datacube is shown in Figure \ref{fig:sl}a with the emitter position overlain as a contour. In the spatrum, the emitter is clearly visible from its [{\sc O\,iii}] and H\,$\beta$, along with weaker [{\sc O\,ii}] and H\,$\alpha$ at $z$\,=\,0.39.

In addition to the nearby emitter, there is a second [{\sc O\,ii}] source at $z$\,=\,0.82 separated by 9.4\,arcsec. We do not consider this in our lensing analysis due to the large separation. There are also a further three clustered [{\sc O\,iii}] emitters, at redshift $\sim$\,0.39. They are offset from the single closely-projected emitter by $>$\,200\,kms$^{-1}$, and so did not originate from a common source. Furthermore the three clustered emitters are unlikely to share a single source, as two are offset in velocity space by $\sim$\,60\,kms$^{-1}$, and they are located 23.5\,arcsec north, and 23\,arcsec west respectively. 


\subsubsection{2MASXJ01260057-0120424}
\label{3C040}

J0126-0120 is a massive ETG ($\sigma_{\rm 6dF}$\,=\,262\,kms$^{-1}$) with a nearby companion separated by 30\,arcsec (the BCG of A0194). The datacube PSF is 0.7\,arcsec. It has a redshift of 0.01824 and has a velocity offset to the cluster of $\Delta v_{\rm cluster}$\,=\,72\,kms$^{-1}$. A0194 has a $R_{500}$ 0.516\,Mpc, and $M_{500}$\,=\,0.40\,$\times$\,10$^{14}$\,M$_{\odot}$ \citep{Piffaretti2011}. The background emitter is separated by 5.4\,arcsec to the south-west (see Figure \ref{fig:sl}b), and has a measured redshift of 0.332. The stellar population of this galaxy was previously studied with Mitchell IFS data, finding radial metallicity gradients in [$\alpha$/Fe] and [Fe/H] \citep[][]{Greene2019}.

\subsubsection{2MASXJ02021730-0107405}
\label{PGC007748}

This system was previously reported in \cite{Smith2018}. 

J0202-0107 (PGC007748) is the BCG of A0295, and located at redshift 0.04276 ($\Delta$v$_{\rm cluster}$\,=\,119\,kms$^{-1}$). The datacube has a PSF of 0.8\,arcsec. This galaxy has a redshift 0.83 emitter separated by 3.5\,arcsec to the north-west, see Figure \ref{fig:sl}c. The strongest emission line is [{\sc O\,iii}]. A0295 is a poor cluster with $M_{500}$\,=\,6.0\,$\times$\,10$^{13}$\,M$_{\odot}$. Therefore the DM will have significantly less of an impact on the lensing mass for this system compared to other, larger clusters, i.e. AS1101, A3395 and A0168. There is a second similarly separated background source 4.2\,arcsec to the north west ($z$\,$=$\,0.83), however as the direction is very similar we will only consider the closer image which will provide stronger constraints.

\subsubsection{2MASXJ02023082-5055539}
\label{J0202}

J0202-5055 is a $z$\,$=$\,0.02148, massive ETG ($\sigma_{\rm 6dF}$\,$=$\,323\,km\,s$^{-1}$). The closest source is separated by only 1.92\,arcsec north-east from the galaxy centre, see Figure \ref{fig:sl}d. This source lies at redshift 0.295, and has a velocity gradient. It was identified from [{\sc O\,ii}], [{\sc O\,iii}] and H\,$\alpha$ emission lines. 

This system also contains another four background emitters, closely projected to the lens. Separated by 3.37\,arcsec to the west is a faint [{\sc O\,ii}] emitter at $z$\,$=$\,1.289. This potentially offers an additional constraint on the mass profile of J0202-5055. A further group of three emitters lies to the north, with a redshift of 0.907 from strong [{\sc O\,ii}] and H\,$\beta$ (the closest separated by 4.68\,arcsec). 

J0202-5055 was observed with SINFONI by \citet{SLC2015} as part of the SNELLS survey, and only the $z$\,$=$\,1.289 emitter is hinted at, close to the frame edge. The new MUSE data uncover a much more complex system, which is potentially a powerful target for future observations, aiming to uncover any counter images to the $z$\,=\,0.29 and $z$\,=\,1.29 sources.

\subsubsection{2MASXJ06273625-5426577}
\label{A3395}

J0627-5426 the BCG of A3395, at $z$\,$=$\,0.04856, with a velocity dispersion of $\sigma$\,=\,276 km\,s$^{-1}$ \citep{Smith2004}. It is offset from the cluster by $\Delta v_{\rm cluster}$\,=\,--\,519\,kms$^{-1}$. We use the combined MUSE-DEEP datacube with a seeing of 1.0\,arcsec. MCXC X-ray data reported $R_{500}$ as 0.930\,Mpc, and $M_{500}$\,=\,2.40\,$\times$\,10$^{14}$\,M$_{\odot}$. Separated by 4.40\,arcsec to the north, we detect the emitter with strong [{\sc O\,ii}], H\,$\beta$, [{\sc O\,iii}] and H\,$\alpha$, at $z$\,=\,0.3971, shown in Figure \ref{fig:sl}e. The is no obvious counterpart to this source in shallow HST observations \citep{Laine2003}.

\subsubsection{2MASXJ12332514-3121462}
\label{J1233}

J1233-3121 is a massive ETG with a 6dFGSv velocity dispersion measured to be $\sigma_{\rm 6dF}$\,=\,348\,km\,s$^{-1}$, and at $z$\,$=$\,0.05194. The background emitter is distant, at 5.14\,arcsec, (still within the predicted 2\,\rein\ for such a high $\sigma$ system), located to the north of J1233-3121, see Figure \ref{fig:sl}f. The emitters spectrum is contaminated by foreground H\,$\alpha$ and [{\sc sii}] emission from the lens galaxy. However, the spectrum is well fit with [{\sc O\,ii}], [{\sc O\,iii}] and H\,$\beta$ at $z$\,$=$\,0.4374, separate from the lens contamination.

\subsubsection{2MASXJ13320334-3146430}
\label{PGC047590}

J1332-3146 has a redshift of $z$\,$=$\,0.04372, and a close star (5.3\,arcsec) located to the south-east. Otherwise its local neighbourhood is sparsely populated, although it lies in a fairly dense region of the Shapley supercluster \citep{Haines2018}. The background emitter is unresolved and separated by 4.24\,arcsec to the west, and is detected via its [{\sc O\,ii}] line at $z$\,=\,0.89, see Figure \ref{fig:sl}g. This system also contains three additional $z$\,$\sim$\,0.89 emitters, all of which also lie significantly more distant from the galaxy centre to the west ($\geq$\,9.34\,arcsec). These will not be included in the analysis due to their respective distance, and as they are not consistent with a multiply-imaged lensing scenario due to a combination of their velocity offsets, image separation and orientation from J1332-3146.

\subsubsection{2MASXJ13522521-3456009}
\label{J1352}

J1352-3456 is an E/S0 galaxy  ($\sigma_{\rm 6dF}$\,=\,341\,km\,s$^{-1}$) with no nearby galaxies of comparable size, at $z$\,$=$\,0.03824. The background emitter is separated by 5.06\,arcsec, to the north-west, and has $z$\,=\,0.1962 from H\,$\alpha$ and [{\sc O\,iii}] emission (see Figure \ref{fig:sl}h).

This system also includes three emitters within 6.9--7.5\,arcsec (See Table \ref{tab:MNELLS}), which are not multiply imaged but are of a similar redshift, all to the south. We make comment on this in Appendix \ref{app:MCE}.

\subsubsection{2MASXJ23135863-4243393}
\label{S1101}

The BCG of AS1101, J2313-4243 lies at $z$\,$=$\,0.05640 ($\Delta$v$_{\rm cluster}$\,=\,--\,480\,kms$^{-1}$), and, as shown in Figure \ref{fig:sl}i, has an emitter separated by 3.6\,arcsec to the south-west. The strongest emission is [{\sc O\,iii}], along with weaker H\,$\beta$ and [{\sc O\,ii}] which is fit at $z$\,=\,0.700. AS1101 has X-ray data from MCXC measuring $R_{500}$ as 0.980\,Mpc, and $M_{500}$\,=\,2.83\,$\times$\,10$^{14}$\,M$_{\odot}$ which is similar to A3395.

\section{Lensing Analysis}
\label{sec:LA}

\subsection{Multiply-imaged cluster-scale systems}
\label{sec:LAMIS}

These cluster-scale strong lenses can only provide relatively poor constraints on the IMF, due to the large image-separation. Here we derive some initial quantities from purely strong lensing analysis for J2357-3445 and J1516+0701 from Sections \ref{sec:A4059}, \ref{sec:A2052}, without attempting to disentangle the DM from the stellar matter.

For each of the two systems we model and optimise the mass profile with {\sc gravlens} to reproduce the image positions, and then estimate the DM contribution for a light profile with a Kroupa or Salpeter IMF. We model both galaxies with a SIE parameterisation, and fix the ellipticity and position angle to the light profile; the normalisation is left to vary. In order to account for extra complexity in the mass model, we include an external shear, with free parameters for the direction and amplitude, which will act as a proxy for the effects of the galaxy DM halo, and external structure in the cluster.

We then compare the lensing masses, within the half-image separation (see Table \ref{tab:CS_lens}), to \MLref\ for Kroupa and Salpeter IMFs (assuming an old stellar population with solar metallicity), to estimate $f_{\rm DM}$. (We measure the luminosity within the Einstein aperture with 2MASS Ks band data for both systems, and higher resolution HST WFPC2 F814W data for J1516+0701.) J2357-3445 requires a dark matter contribution of 76 per cent even for the higher stellar mass Salpeter IMF. Likewise J1516+0701 requires a 78 per cent DM contribution to have a stellar population comparable with a Salpeter IMF. Hence for any plausible IMF, the lensing mass is dominated by the DM rather than the stars. This highlights the importance of finding lenses for which the \rein\ probes the most central most stellar dense regions of the lens. The higher resolution HST WFPC2 F814W data, measured \ML is found to be in close agreement with that measured from the lower quality 2MASS data (see Table \ref{tab:CS_lens}).

\begin{table*}
	\caption{The lensing analysis of the two cluster-scale lenses. The quoted masses are {\it total} lensing masses, including dark matter. The luminosity is Ks band 2MASS for J2357-3445, and WFPC2 F814W and 2MASS Ks for J1516+0701 The uncertainty is 0.1\,arcsec for $R_{\rm ap}$. This follows through into the aperture mass as a 10 per cent uncertainty, and into the aperture luminosity as 2 per cent. We estimate that the uncertainty in the dark matter fraction is $\sim$\,5 per cent.}
	
	\label{tab:CS_lens}
\begin{tabular}{|c|c|c|c|c|c|c|c|c|}
	\hline Name & band & $R_{\rm ap}$ (arcsec) & $M_{\rm ap}$ (10$^{10}$ M$_{\odot}$) & $L_{\rm ap}$ (10$^{10}$ L$_{\odot}$) & \MLref & mass-excess ($\alpha$) & $f_{\rm DM}^{\rm Kroupa}$ & $f_{\rm DM}^{\rm Salp}$ \\ 
	\hline 
	J2357-3445 & Ks & 8.5 & 179 & 28.34 & 0.96 & 6.56 & 85 & 76 \\  
	J1516+0701  & Ks & 9.7 & 165 & 24.22 & 0.97 & 7.02 & 86 & 78 \\ 
	& F814W &  &  & 8.84 & 2.66 & 6.99 & 86 & 78 \\ 
	\hline 
\end{tabular} 
\end{table*}

\subsection{Single-image galaxy-scale systems}
\label{sec:LASIS}

Whilst a multiply-image system provides the strongest constraints on the IMF, there is information stored in systems with only a single, close-projected emitter. In these cases we can constrain the \ML of the `lens' as having to be consistent with mass profiles which do {\it not} produce a detectable counter-image. In turn, this translates to a maximal mass-to-light ratio excess parameter ($\alpha$), and hence adds further constraints on the IMF in ETGs.

Here we present the analysis of the nine identified single-imaged close-projected systems. In order to be self-consistent we try to select a common source of K$_s$-band imaging for all candidates. The highest resolution data for this purpose is from the VISTA Hemisphere Survey \citep{McMahon2013}. Only J0627-5426 is not covered by VISTA. For this target we take the poorer resolution imaging from the 2MASS survey \citep{Jarrett2000}. The average image PSF FWHM for the VISTA data is 1\,arcsec, and for 2MASS is 3\,arcsec. We derive all of our quantities in the K$_s$-band, using the Vega solar absolute magnitude quoted by {\sc ezgal} \citep{Mancone2012} for the 2MASS K$_s$-band in their filter list (M$^{\odot}_{\rm K_s}$\,=\,3.295).

We perform our lensing analysis with {\sc gravlens} \citep{Keeton2001}, using pixelised mass maps derived from the light profile of each galaxy. The mass profile is well traced by the light at scales comparable with, or smaller than, the \reff. The central 7\,arcsec of each galaxy is fit with a single S{\'e}rsic profile using {\sc galfit} \citep{Peng2010}, fixing only the sky. Then the de-convolved model is converted into a pixelised mass map (with a fiducial scaling of \ML\,=\,1). The mass maps are input to {\sc gravlens} and a range of shear ($\gamma$; 0--0.2) and shear position angles ($\theta$; 0--180 deg) are applied for each scaling of \ML\ (\ML; 0.5--3.0). We select the rms for each Cartesian shear component to be $s$\,=\,0.05 as these are galaxy lenses without significant nearby mass distributions.  



\begin{figure*}
\centering
\includegraphics[width=1.0\linewidth]{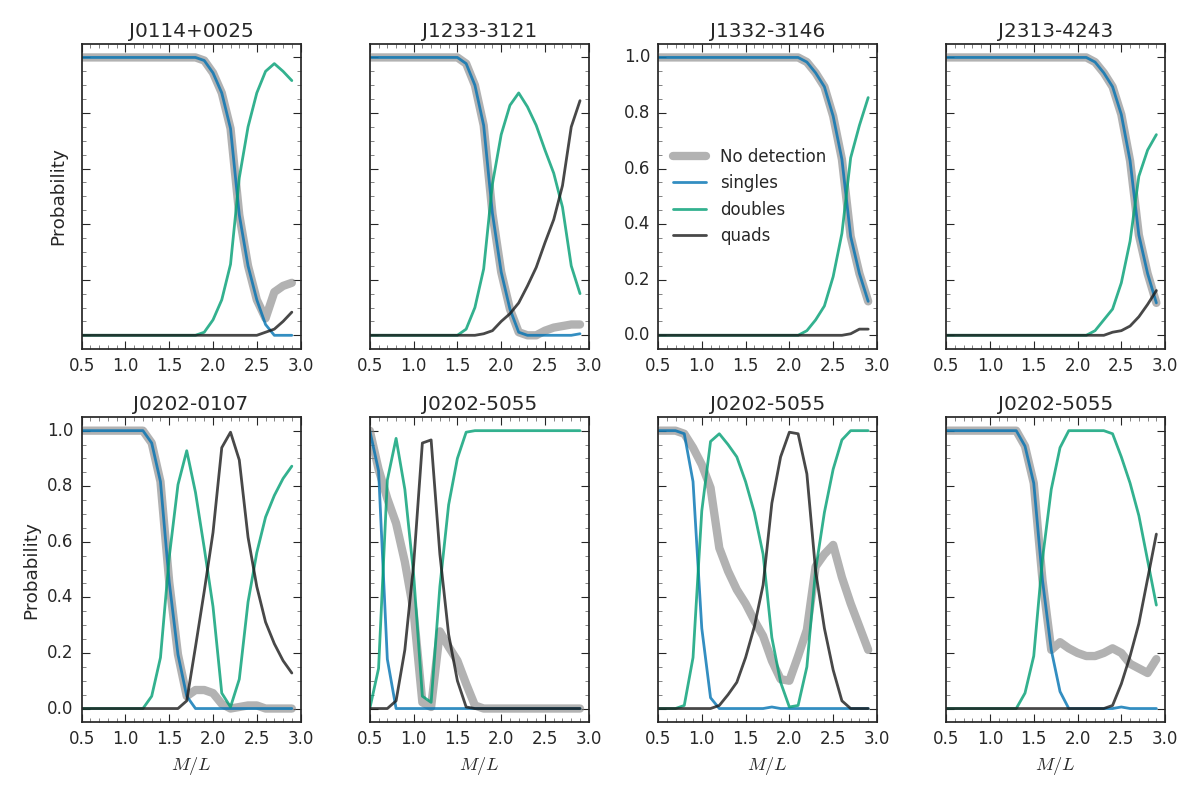}
\caption{Upper-limit results for six galaxies with a single closely-projected image, plotted as probability against K$_s$-band mass-to-light ratio (\ML). The grey lines denote the probability of {\it not} detecting a counter-image in each case. The blue/green/black lines are the fraction of intrinsic singles/doubles/quads within each \ML\ bin as the requirement to reproduce the observed image is met. These range from producing multiple images only at the highest of \ML, J1332-3146, to J0202-5055 which for even low \ML\ there are many more intrinsically multiply-imaged systems. The systems J0202-0107 and J0202-5055 are constrained to have the lowest \ML\ in the sample.}
\label{fig:single_lensing_results}
\end{figure*}

The mass maps are used to generate a grid of lensing models constrained by the position of the observed emitter. Then the number of detectable counter-images to the observed image is extracted. Following the framework outlined in \citet{Smith2018}, we estimate the probability that the system only has a single detected image, for each trial value of \ML. The results are shown in Figure \ref{fig:single_lensing_results}, with the {\it intrinsic} multiplicity curves showing the lensing regimes as we move from low to high \ML.

The curves display the regimes in which the background emitter is intrinsically singly--imaged, then doubly--imaged, quadruply--imaged and then returns to doubly-imaged as the source position required to recreate the observed image moves across the caustic lines on the source plane. Additionally we show the likelihood of there being no {\it detectable} counter image for each \ML\ bin (i.e. including the systems which are intrinsically multiply-imaged, but where the counter-images are expected to be too faint to detect). We establish the S/N limits for an undetectable counter-image by re-inserting the detected source at random positions close to the foreground galaxy with different flux scalings. Each image is then visually inspected to determine whether the source would be recovered.

As all of our quantities are derived in the K$_s$-band, we can interpret the results with regards to a different choice of IMF via the mass-excess parameter, $\alpha$ = (\ML)/\MLref. We compare the measured \ML\ to a plausible range of \MLref\ under the assumption of an old, metal rich population typical of early-type galaxies, and adopting a Kroupa IMF. Using the \citet[][hereafter C09]{Conroy2009} models accessed with {\sc ezgal}, for populations of metallicity 1\,--\,1.5\,Z$_{\odot}$, and formation age 10--12\,Gyrs, the sample galaxies have K-band \MLref\ in the range 0.9\,$<$\MLref\,$<$\,1.1. 

Our choice of the C09 models leads to a subtle difference between this work and \citet{Smith2018}, who used models from \citet[][]{Maraston2005}. The \MLref\ for a Salpeter IMF tabulated by C09 and M05 agrees to within a few per cent in the 2MASS Ks-band, (for old populations with solar metallicity) \citep[see][]{Mancone2012}. However, due to different treatments of the low-mass stars and the intrinsic uncertainty in the passive luminosity evolution of a galaxy, the ratio of $\alpha_{\rm Kroupa}$ and $\alpha_{\rm Salpeter}$ is different by $\sim$\,8 per cent. Therefore in this paper for a result in agreement with a Salpeter IMF, $\alpha_{\rm Kroupa}$\,$\simeq$\,1.64, instead of $\simeq$\,1.52. Secondly, and key, is the adopted solar absolute magnitude. The K-band tabulated by M05 relates to the Johnson-Cousins K, used in \citet[][hereafter M98]{Maraston1998}, prior to the advent of the `short' K (i.e. Ks) filter which dealt better with zero point issues due to H$_2$O in the atmosphere \citep[see][]{Bessell2005}. Hence, the M98 M$^{\odot}_{\rm K}$\,=\,3.41, is 0.115 mags offset from M$^{\odot}_{\rm Ks}$\,=\,3.295, and this leads to a 11 per cent decrease in the derived luminosity when the correct value is used.

For several galaxies in this sample, the background emitter is not multiply-imaged within the full range of \ML\, and are therefore excluded from any further analysis (J0126-0120, J0627-5426 and J1352-3456). In addition, from those presented in Figure \ref{fig:single_lensing_results}, J2313-4243, J1332-3146, J0114+0025 and J1233-3121 are each single-imaged at masses larger than predicted by a Salpeter IMF (\ML\,=\,1.6). Therefore we do not make further comment on these systems in this Section. However, the systems are included for investigating the ETG population in an ensemble sense, in Section \ref{sec:TLPop}.



We will now convert from \ML, to $\alpha$ for the two galaxies with the strongest constraints (J0202-0107, J0202-5055). This conversion uses \MLref\ from C09, for an old population, typical for low-$z$ ETGs. This therefore sets a MW-like IMF to have $\alpha$\,=\,1.0, a Salpeter IMF to have $\alpha$\,=\,1.64 and we define a "heavyweight" IMF with $\alpha$\,=\,2.0. We do not model the contribution from DM within the lens galaxies, therefore the $\alpha$ measurements are slightly overestimated. The current low-$z$ lenses have a correction of about 20 per cent \citep{SLC2015,Collier2018a}. A system which is expected to form multiple images with a relatively light (i.e. Kroupa) IMF, is a promising candidate for deeper follow-up, as the observational depth is likely to be the limiting factor. In the following subsections, \ref{subsec:PGC07748} and \ref{subsec:J0202-5055} we will describe J0202-0107, J0202-5055 respectively in reference to the IMF.



\begin{figure*}
	\centering
	\includegraphics[width=1.0\linewidth]{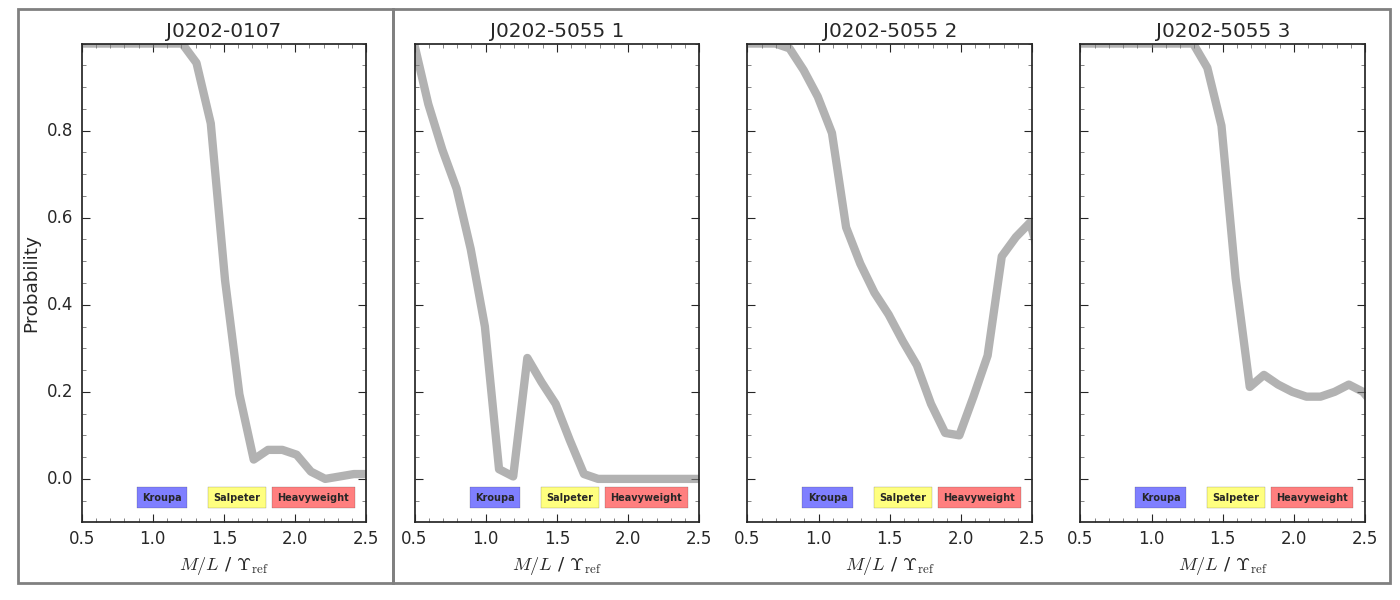}
	\caption{The constraints for the upper-limit lensing analysis converted from K-band $M/L$ to $\alpha$. The grey lines show the probability of the source having no {\it detectable} counter-image. Marked along the horizontal axis are approximate bounds for different choices of IMF. The panels are labelled for the galaxy they relate to, {\it left:} J0202-0107, {\it middle left/ middle right/ right:} J0202-5055 inner source, outer source, distant source. }
	\label{fig:single_lensing_alpha}
\end{figure*}

\subsubsection{J0202-0107}
\label{subsec:PGC07748}


J0202-0107 transitions from intrinsically singly-- to multiply--imaged for \ML\ between 1.25--1.75, see Figure \ref{fig:single_lensing_results} (top, centre right, blue and green tracks). As this source is bright, the probability of a non-detection (thick grey) follows very closely the intrinsically single-imaged (blue) track. After converting from \ML\ to $\alpha$ it is clear for any IMF heavier than Salpeter the emitter must be intrinsically multiply-imaged (Figure \ref{fig:single_lensing_alpha}).

If we compare our 50 per cent probability cut off, to that reported in \citet*{Smith2018}, this result is 20 per cent larger. The revised solar absolute magnitude contributes to a 12 per cent increase. The remaining difference must originate from other modelling uncertainty. In the earlier paper we fitted a de Vaucouleurs profile ($n$\,=\,4) to a Pan-STARRS $y$-band image and scaled the resulting model to a 5\,arcsec aperture 2MASS K$_s$ flux. We now instead fit directly to VISTA K$_s$-band imaging, and allow a free S{\'e}rsic parameter. This leads to a difference of 0.05\,mags between the apparent magnitude here using the de-convolved {\sc galfit} model, compared to the psf-corrected measurement from 2MASS (due to the much larger 2MASS PSF). Finally, our cosmology leads to a 3 per cent change, with an increased H$_0$ value. Therefore, the changes in how the lens light is modelled, and the cosmology account for the difference relative to the earlier work.




\subsubsection{J0202-5055}
\label{subsec:J0202-5055}

The nearest three emitters in the J0202-5055 system are displayed in Figure \ref{fig:single_lensing_results}. We do not consider the effect of multiple-plane lensing for any of the sources. The source with the smallest separation is also at the lowest redshift, so there is no multi-plane effect in this case. In principal there is an impact on the other two sources from those preceding in redshift, but the effects are likely negligible because the lensed galaxies are likely to be very low-mass.


The inner source is separated from the galaxy centre by 1.92\,arcsec with $z$\,$=$\,0.29, and the outer source is separated by 3.4\,arcsec, at $z$\,$=$\,1.29. The inner source is expected to be {\it intrinsically} multiply-imaged even for stellar populations described by \ML\,$=$\,1. As the source is faint, and the probability for a non-detection is low in the quad regime, and a small tail at \ML\,=\,1.2--1.5. Hence the depth of our observation is the limiting factor for a non-detection of a counter-image for this source. The second source is more distant, and hence can be intrinsically single-imaged for more massive systems up to \ML\,$\simeq$\,1.1--1.3.



As with J0202-0107, we convert from \ML\ to $\alpha$ in order to investigate this system with respect to the IMF (See Figure \ref{fig:single_lensing_alpha}). If the foreground galaxy formed with an IMF consistent with Kroupa, the first source should be multiply-imaged, but the second and third sources are likely not. If we consider a Salpeter IMF, or an $\alpha$ consistent with that predicted from Atlas3D \citep[$\sim$\,1.5,][]{Cappellari2013} the first and second source are intrinsically multiply-imaged. A galaxy forming with a "heavyweight" IMF may even produce counter-images for all three sources.


J0202-5055 offers a promising system for further observations, as for any reasonable IMF parametrisation a deeper observation is likely to unveil a counter-image to the innermost source. Analysis can then constrain $\alpha$, using the same technique as the other low-$z$ systems \citep[i.e. J0403-0239 in][]{Collier2018b}. In addition, the second faint source, separated by 3.4\,arcsec, offers a potential double source-plane if the lens IMF is even modestly heavier than Kroupa (including DM).

\subsubsection{Summary}

Out of the nine apparently single-imaged systems presented in Figure \ref{fig:sl}, analysis showed two systems which likely have faint counter images for even a low \ML\ lens (J0202-0107, J0202-5055, see Figure \ref{fig:single_lensing_results}). However, more detailed analysis shows that J0202-0107 cannot rule out having formed with a moderately heavy IMF ($\alpha$; 1.25--1.75). This range is consistent with the $\alpha$-vs-$\sigma$ relation measured with stellar dynamics by \citet{Cappellari2013}, which predicts $\alpha$\,=\,1.5\,$\pm$\,0.3 for $\sigma$\,=\,264\,kms$^{-1}$. For J0202-5055, if the  lens forms with an IMF consistent with the $\alpha$-vs-$\sigma$ relation, then a double source plane lens will be discovered with deeper observations. If confirmed by future observations, J0202-5055 will be one of only a handful of known double source-plane systems.


\subsection{The Lens Population}
\label{sec:TLPop}

The {\it intrinsic} distribution of the IMF within ETGs can be inferred from the four confirmed low-$z$ lenses \citep[taking values for SNL-0, SNL-1, SNL-2, J0403-0239 from][]{Newman2017,Collier2018a,Collier2018b}. We have shown in \citet{Collier2018b} that for a normal distribution with a flat prior on the mean and marginalising over the intrinsic scatter, $\langle \alpha \rangle$\,=\,1.09\,$\pm$\,0.08. Alternatively, by marginalising over the mean, the 90 per cent upper limit on the standard deviation of $\alpha$ is 0.32 (29 per cent). However, if there is a sizeable scatter in the IMF within the ETG population, then the lensing systems are likely to be biased towards those with the highest mass. This was found by \citet{Sonnenfeld2019}, who compared the SLACS lens population to weak lensing of non-strong-lensing ETGs. Therefore combining the `Upper limits' and the confirmed lenses may better constrain the {\it ensemble} population \citep[see also][]{Shu2015}. Here, we use seven of the nine upper limit systems presented in this paper, (due to the lack of constraint even at the top of our \ML\ range, we do not include J0627-5426 and J1352-3456). We add the two systems analysed in \citet{Smith2018} from other surveys (SNL-4 and J0728+4005) along with the four confirmed lenses.

\begin{figure*}
	\centering
	\includegraphics[width=1.0\linewidth]{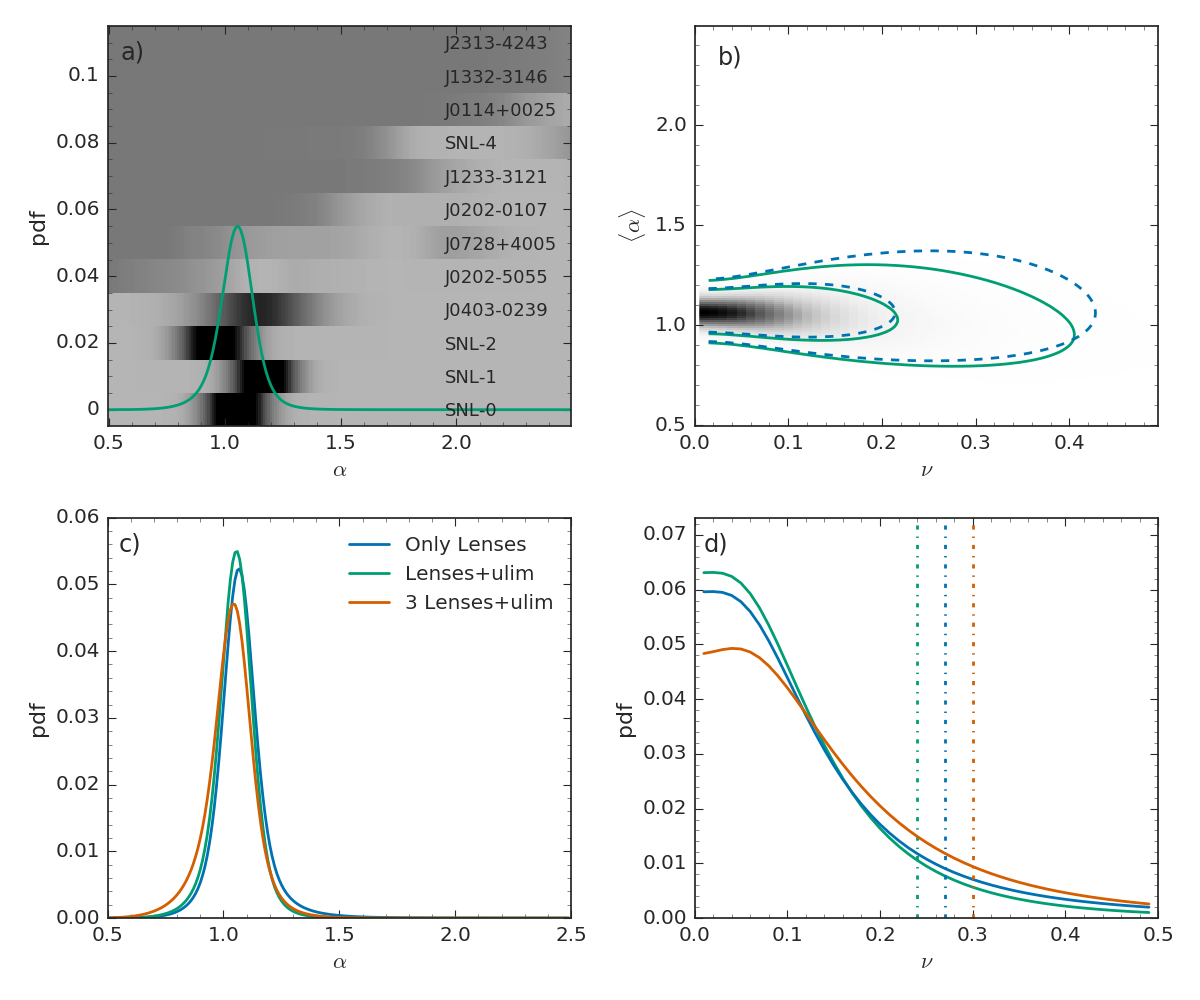}
	\caption{The distribution of $\alpha$ for the population of both lenses and single-imaged systems. Values calculated with just the four lenses are in blue, the four lenses combined with the upper limit analysis is green, and removing J0403-0239, leaving 3 lenses and the upper limits is displayed in orange. Panel a) The lenses and upper limits intrinsic population in $\alpha$. The shaded background regions show the probability of a given alpha for the lenses (lowest four), or upper limit analysis. Panel d) Contours showing the distribution in $\langle \alpha \rangle$ and $\nu$ (the intrinsic scatter) space with and without the upper limits. The twist to lower $\alpha$ at higher $\nu$ can clearly be seen. Panel c) The intrinsic distribution in $\alpha$ as we change the sample. Removing a the lens J0403-0239 has the largest effect, and shifts the distribution to favour lower $\alpha$. Adding the upper-limits shifts the peak of the distribution to lower $\alpha$, compared to just the lenses. This is due to the significant effect of J0202-3055. Panel d) The predicted intrinsic scatter of the distribution in base $e$, with dashed lines at the 90 per cent confidence interval for each case. As can be seen, adding the upper-limits favours a population with a smaller scatter, than just the lenses. Removing J0403-0239, and including the upper-limits increases the intrinsic scatter significantly (30 per cent).}
	\label{fig:alpha_distr}
\end{figure*}

We model the intrinsic population as a lognormal (base $e$) distribution in $\alpha$, described with a mean, $\langle \alpha \rangle$ and an intrinsic scatter $\nu$, as currently there are too few systems to constrain the shape of the distribution. Modelling $\alpha$ with a normal distribution does not significantly affect the results. We set flat priors on $\langle \alpha \rangle$ and $\nu$, between [0,\,2.5] and [0,\,0.5] respectively. We derive the probability of each $\alpha$ value, given either a confirmed lens, or the upper-limit analysis, see the shaded regions in Figure \ref{fig:alpha_distr}a. Systems alike to J0202-5055, which have tight upper-limits on $\alpha$ {\it comparable} with those measured from confirmed lenses, offer the most information. 


For each confirmed lens, we estimate the likelihood of drawing $\alpha_{\rm meas}$ from the intrinsic population, $P$($\alpha$|$\langle \alpha \rangle$,$\nu$) with a broadening on $\nu$ from the uncertainty in the measurement. A grid based exploration of $\langle \alpha \rangle$, and $\nu$ with step size 0.01, produces the dashed contours shown in Figure \ref{fig:alpha_distr}b. The intrinsic distribution marginalised over $\nu$ has a $\langle \alpha \rangle$\,=\,1.07\,$\pm$\,0.09, shown in Figure \ref{fig:alpha_distr}c, which is consistent with the result of $\langle \alpha \rangle$\,=\,1.09\,$\pm$\,0.08 from \citet{Collier2018b} for the same lenses.

To include the upper limits a slightly more complex approach is required. The probability of a given $\langle \alpha \rangle$, $\nu$ combination (P$_i$($\langle \alpha \rangle$,$\nu$)) is now given by; 
\begin{equation}
    P_i(\langle \alpha \rangle,\nu)\,=\, \int U(\alpha_i) P(\alpha_i | \langle \alpha \rangle,\nu) d\alpha_i
\end{equation}
where $U(\alpha_i)$ is the likelihood for a galaxy to have a given $\alpha$ (in the range [0,\,3.0]), from the `upper limit' analysis. This is related to $U$(\ML) by a convolution with a Gaussian uncertainty contributed by \MLref. Including this likelihood allows us to marginalise over the unknown true value of $\alpha$ for each galaxy.


The distribution from the upper limits alone is strongly skewed towards low $\langle \alpha \rangle$. Therefore combining this with the confirmed lenses, skews the overall distribution towards lower $\langle \alpha \rangle$ (see Figure \ref{fig:alpha_distr}c). It also reduces the scatter $\nu$. The combined distribution has a mean of 1.06\,$\pm$\,0.08, shown in Figure \ref{fig:alpha_distr}a. The 90 per cent confidence upper limit on the standard deviation of $\alpha$ is 0.24 (20 per cent). This is a nine per cent decrease compared to using only the four confirmed lenses. 

We consider what finding additional single-imaged galaxies might do to the distribution by testing the addition of hypothetical systems consistent with those presented in this paper. If we add new fake systems consistent with J2313-4243, we find that these do not change $\langle \alpha \rangle$ or $\nu$. However, adding four hypothetical systems alike to J0202-0107, where the 20 per cent probability for a non-detected counter image lies at $\alpha$\,=\,1.5, reduces the inferred intrinsic scatter in the population, without changing $\langle \alpha \rangle$. Finally, additional hypothetical systems consistent with J0202-5055 and J0728+4005 will {\it increase} $\nu$ and {\it reduce} $\langle \alpha \rangle$, as the upper-limits suggests that they formed with IMFs lighter than $\langle \alpha \rangle$ predicted from only the confirmed lenses. 


Our analysis demonstrates that the strong-lens systems favour a comparably heavier IMF than those systems which include only a single emitter. Therefore although our population of strong lenses are offset to \ML\ larger than a Kroupa IMF (by 8 per cent), these are biased towards higher $\alpha$, where as the distribution which will have a smaller mean. Hence the {\it ensemble} population lies in closer agreement with the Kroupa IMF. Our distribution of $\langle \alpha \rangle$\,=\,1.06\,$\pm$\,0.08 is within 3$\sigma$ of similar attempts to combine strong-lensing constraints with other independent techniques \citep[e.g.][]{Sonnenfeld2019}, who found $\alpha$\,=\,0.80\,$\pm$\,0.11.

\section{The MNELLS search efficiency}
\label{sec:RoD}

In this Section we will assess the MNELLS programme technique. In Section \ref{sec:DetThresh} we test the flux detection threshold as a function of wavelength, and then of distance from the centre of the target galaxy. In Section \ref{sec:Ndenemitters}, we determine the number density of background emitters in our dataset. Then in Section \ref{sec:Disc} we discuss potential modifications to the observing strategy, and contrast our technique to the MaNGA survey.

\subsection{Detection threshold with artificial point sources}
\label{sec:DetThresh}


The flux detection limit for a background source is the key parameter for assessing the efficiency of the MNELLS observing setup. This will test whether the true observed depth matches our expectation based on other surveys and inform plans for future observing campaigns. 


We compute the recovery fractions by injecting fake point sources, convolved to a representative psf, into a real datacube. There is scatter in the sky background from galaxy to galaxy, so we select J2318-1023 which contains few other bright emitters and has noise properties representative of the median of our sample. Each source is modelled with a gaussian with a FWHM of 5\,\AA\, and a spatial FWHM matched to the datacube seeing of 0.8\,arcsec, typical for an emission line within our survey. Sixteen sources are injected into to the datacube as a 4\,$\times$\,4 source grid, with a subpixel scatter applied to each position in both the spatial and spectral dimensions. For each wavelength and flux we applied nine dithers to the grid. The sources are then scaled to total fluxes ranging from 10$^{-15.25}$ to 10$^{-18.5}$\,erg\,s$^{-1}$\,cm$^{-2}$, in steps of 0.25 dex.



We perform this test for four wavelength channels which are representative of the typical noise situations for the background emitters. The channels are chosen as 5000\,\AA, 6861.25\,\AA, 7100\,\AA, 7242.5\,\AA, which covers the blue, less sensitive region of the datacube (5000\,\AA), a mid-way low sky noise region (7100\,\AA), and then two regions of the datacube close to sky lines. These are selected to match the wavelengths used in equivalent analysis by \citet{Herenz2017}. For each cube with simulated sources the full processing and emission line detection is carried out, as described in Section \ref{sec:gseld}, and the recovery fractions are measured.



\begin{figure}
	\centering
	\includegraphics[width=\linewidth]{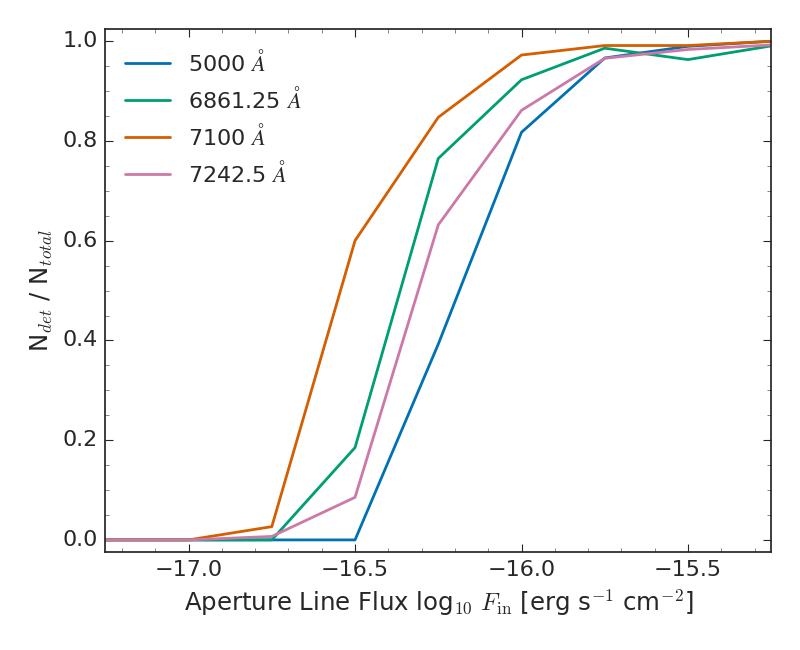}
	\caption{Recovery fraction detected/injected, from the insertion and detection of fake point sources for four different wavelengths within a single MUSE datacube. The detection process follows that used for the targeted observations.}
	\label{fig:senslim}
\end{figure}

The most sensitive of these channels is the 7100\,\AA\ channel, with a 90 per cent threshold of $\sim$10$^{-16.15}$ erg\,s$^{-1}$\,cm$^{-2}$. In contrast, the bluest channel (5000\,\AA) has a significantly lower sensitivity, $\sim$0.4 dex offset from 7100\,\AA\, see Figure \ref{fig:senslim}. This is likely due to a high lunar continuum, i.e. a high fraction of lunar illumination (FLI), which ranges up to 0.9, during the time of our observations. Note that in practice, the impact of a low detection threshold in blue channels could be reduced as the emission lines detected in the blue are mainly low-$z$ [{\sc O\,ii}] emitters. These may still be identified from strong [{\sc O\,iii}] and H\,$\alpha$ lines, which will be present in lower background redder channels. The only lines with no strong counterparts in the cleaner spectrum are Ly-$\alpha$ emitters. 

For comparison to these results, we have performed the same fake source injection into a MUSE-Wide datacube. We reach a threshold $\sim$\,0.5 dex fainter (at 7100\,\AA) with this data. The difference originates from the sky background due to their observations taking place in dark time. After the ETG is filtered, the MNELLS background is on average three times higher, with the most significant difference seen at 5000\,\AA\ (due to the high FLI), and the weakest at 7242.5\,\AA\ where the noise is instead dominated by a bright sky line.

\begin{figure}
	\centering
	\includegraphics[width=\linewidth]{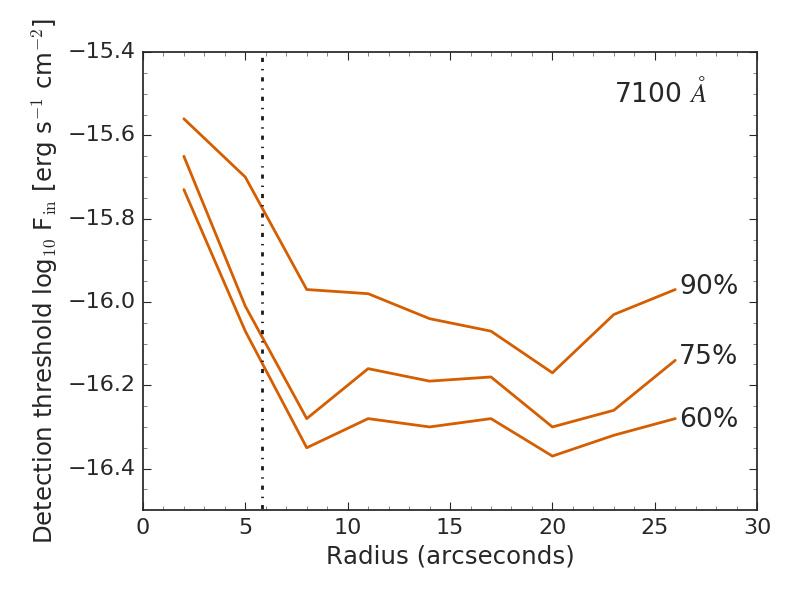}
	\caption{The detection threshold radially binned for fake sources injected into a real datacube. This is for only the `best' wavelength channel from Figure \ref{fig:senslim} is taken (7100\AA). We show the detection limit within each bin for changing recovery fraction. There is a large drop in the threshold for the centremost radii due to the subtraction of the foreground candidate `lens'. The typical predicted 2\rein\ is marked with a dashed line.}
	\label{fig:senslimrad}
\end{figure}

Our search is designed to target sources which are close to the centre of a foreground galaxy. Hence we incur a much larger contribution to the noise than just the sky background at small lens-source angular separation. Figure \ref{fig:senslimrad} shows the detection limit for the 7100\,\AA\ channel as a function of radius. Within 6\,arcsec of the galaxy centre ($\sim$\,2\,\rein), the flux threshold is typically 0.35 dex brighter than the full FoV. Beyond a radius of 10\,arcsec the sensitivity shows no strong radial dependence.



\subsection{Number density of background emitters}
\label{sec:Ndenemitters}

In this subsection we consider the number density of background emitters actually detected within our targeted observations. We also compare to the MUSE-Wide for our measured flux threshold. As described in Section \ref{sec:MNELLS}, the MNELLS observations are shallow exposures taken in poor seeing conditions. In our calculations we only include the fourteen fields which have the full exposure time; in total these fields have 164 detected emitters. The average full depth area within a combined datacube is 0.92 square arcmin, therefore the median number density of emitters within our full depth observations is 12.7 per square arcmin; the field-to-field standard deviation is 5.5.

The full depth MUSE-wide survey reaches a number density of 37.4 emitters per square arcmin. In comparison with MNELLS, each observation is 20 per cent longer (a 1\,hr exposure), in conditions with a lower sky background (observed in dark/grey time instead of grey/bright). So, using their figure 13, and with our cut off at a depth of 10$^{-16.1}$ erg\,s$^{-1}$\,cm$^{-2}$ (for 90 per cent to be detected), we would predict 6.3 emitters per square arcmin. The number density from our MUSE data (12.7 per square arcmin) is actually twice this. The detection threshold which aligns these two values is a flux limit of 10$^{-16.5}$ erg\,s$^{-1}$\,cm$^{-2}$. From Figure \ref{fig:senslim}, this suggests we should adopt a 60 per cent detection limit to align our numbers with the MUSE-Wide.

For the central 6\,arcsec where the detection threshold for 60 per cent recovery (to match the achieved number density of emitters) is 10$^{-16.15}$ erg\,s$^{-1}$\,cm$^{-2}$, the MUSE-wide predicted number count is 6.3 per square arcmin (an 80 per cent decrease). For the average targeted galaxy, the strong lensing regime encompasses 23 square arcsec ($\pi$\rein$^{2}$; \rein\,=\,2.9\,arcsec). The resulting probability for a lens galaxy having a detectable multiply-imaged source is 0.04. So we expect one in 25 observed galaxies will exhibit a multiply-imaged background source. Note that the counter-images may be fainter than the detection threshold.




\subsection{Considerations for future MNELLS-like surveys}
\label{sec:Disc}


The MNELLS project observational constraints (e.g. no moon constraints) were selected to maximise the number of executed observations by taking advantage of under-used observatory conditions, under the assumption that the ETG will dominate the background. We predict the number of lenses to be one in every twenty-five full depth observations, and in calculating this we uncovered a few key contributions affecting our detection limits which need to be considered for future targeted surveys.

The background noise from the lunar continuum, and sky brightness has clearly affected our detection limits. The variable observing conditions (specifically the FLI and sky background) may provide a simple explanation for the origin of the large field-to-field scatter in the number of detected emitters, independent of the lens light. In addition the subtraction of the ETG also negatively impacts the detection threshold. Hence, these two effects (higher background, and lens subtraction) reduce our detection threshold compared to the MUSE-Wide survey by almost 0.5 dex. The centre of the datacube is less sensitive to faint emitters, which cannot be avoided, but at shorter wavelengths we are also offset by 0.4 dex from channels less affected by the lunar continuum. For future surveys, as a trade off for the number of observations, a tighter constraint on the moon illumination may be considered to reduce the sky background.

In addition to altering the observing conditions, we can consider an alternative selection method, which may improve the likelihood of a lens discovery. Our current selection criteria (Section \ref{sec:MNELLS}) uses a single-fibre velocity dispersion as a proxy for stellar mass, and hence \rein. Instead we could directly measure \rein\ from a mass-follows-light profile, (such as those used in our upper-limit lensing calculations), and select galaxies to maximise the lensing cross-section. However, using the light profile measured \rein\ as a part of the selection process introduces additional computation, which will require additional cuts in parameter space to be efficient. Instead the current technique is considerably simpler, and the \rein\ scales with the velocity dispersion \citep{Bolton2008}.


An alternative to the "targeted" blind search methods like MNELLS is to exploit large IFU surveys. For example the Mapping Nearby Galaxies at APO (MaNGA) survey \citep{Bundy2015} has a sample median redshift of $\langle\,z\,\rangle$\,=\,0.05 (100 $<$\,$\sigma$\,$<$ 400 kms$^{-1}$), which is comparable to MNELLS. The full MaNGA survey will take IFU observations of over 10\,000 ETGs and hence will discover a number of low-$z$ strong gravitational lenses. To date, \citet{Talbot2018} worked with a sample of 2812 ETGs, and so far has produced 2--6 candidate lenses.

Although MaNGA's median redshift is comparable to that of MNELLS, it aims to observe a broad range of galaxy sizes, including a high fraction of low-$\sigma$ ETGs. The larger (higher-$\sigma$) ETGs typically lie more distant than their median redshift, so that the same fraction of \reff\ is imaged within the FoV \citep[See Fig.8,][]{Bundy2015}. This has two implications: First, the low-$z$, low-$\sigma$ galaxies are less likely to be gravitational strong lenses so the number of discovered lenses is likely reduced. Secondly, the higher $\sigma$ galaxies which are more likely to be lenses lie and higher redshift and are hence subject to larger DM uncertainties. Such lenses, \citep[e.g. J1701+3722, ][$z$\,=\,0.12]{Smith2017,Smith2019}, are complimentary to SNELLS/MNELLS nearby lenses, but not identical. However, lenses within this redshift range may offer an avenue to address the differences between the SLACS and SNELLS/MNELLS conflicting samples.

\section{Conclusion}
\label{sec:Conc}

We have presented the first results from the ongoing MNELLS survey, and a complementary archival sample, searching for nearby strong-lensing early-type galaxies. We have observed sixteen new targets (fourteen to full depth), and selected thirty-six comparable galaxies which were publicly available from the ESO archive at the end of February 2019. We search within each datacube for background sources which are multiply-imaged or, single-imaged but closely-projected to the `lens'. We processed each pipeline reduced datacube to filter out the foreground galaxy, and then search the residuals for emission lines with an automated detection process. We then manually verify the detections, and find that human interpretation is key to reliable identification.

From the total sample, we have discovered one new galaxy-scale strong lens, J0403-0239, which is consistent with a MW-like IMF, $\alpha$\,=\,1.15\,$\pm$\,0.17 \citep*[see][]{Collier2018b}, for an assumed old stellar population. The newly acquired, high S/N, FORS2 spectrum of J0403-0239 will be used to precisely constrain the age of the stellar population and hence $\alpha$. A further two new low-$z$ cluster-scale lenses were discovered in which the lensing mass is dominated by DM for any plausible IMF.


We have discovered nine galaxies with closely-projected but single-imaged sources ($r$\,$<$\,6\,arcsec), which we use to augment the constraints on the IMF via our `Upper-limit lensing' technique \citep*{Smith2018}. Of these systems, J0202-5055 shows a hint of having a counter-image to its innermost source, and hence could become the fifth low-$z$ lens with additional observations. If the inner image has no counter-image the configuration implies a system which is likely consistent with a MW-like or lighter IMF. Of the other eight, only J0202-0107 favours a Salpeter or lighter IMF.

Combining the four confirmed low-$z$ lenses with nine closely-projected but single-imaged systems from this paper, and two taken from \citet[SNL-4 and J0728+4005][]{Smith2018} we infer the mean and intrinsic scatter of the ETG population, assuming a lognormal distribution for $\alpha$. The population has $\langle \alpha \rangle$\,=\,1.06\,$\pm$\,0.08, and a 90 per cent confidence standard deviation of $\alpha$ of 0.24. This is less than the $\langle \alpha \rangle$\,=\,1.09\,$\pm$\,0.08, and $\sigma$\,$<$\,0.32 from only the confirmed lenses \citep{Collier2018b}. This implies that there is a higher likelihood of discovering strong-lensing around massive galaxies which have higher stellar mass (i.e. more compact) due to their increased surface mass-density. Therefore, although SNELLS, and MNELLS lenses on average measure an $\alpha$ a few percent larger than MW-like, the total population lies close to a MW-like IMF.

We tested our detection flux thresholds within the MNELLS datacubes, and predict that we reach a flux limit of 10$^{-16.5}$ erg\,s$^{-1}$\,cm$^{-2}$. Our central sensitivity is 0.35 dex lower due to the foreground galaxy continuum. From the MUSE-wide results, we predict that one in twenty-five targets will be multiply-imaged strong lenses. So far with fourteen targets we have two closely-projected but single-imaged sources and no strong lenses (our reported strong-lensing galaxies are from the archival sample), which is consistent within Poisson statistics of these predictions.

This paper has presented the first of a series of MNELLS observations, with 21 more galaxies observed in P103 and P104 as of Feb 2020. Additionally, FOCAS on Subaru \citep{Ozaki2014} offers an alternative to MUSE, and we are extending our survey technique to the northern hemisphere. From this dataset J0202-5055 is a promising candidate for future follow-up observations as it appears a highly likely to be a multiply-imaged strong-lens for any typical IMF, and has the potential for multiple source planes. With this large array of incoming data, the likelihood of discovering new low-$z$ strong lensing galaxies in the forthcoming years is high.


\section*{Acknowledgements}


W. Collier was supported by an STFC studentship (ST/N50404X/1). RJS and JRL are supported by the STFC Durham Astronomy Consolidated Grant (ST/P000541/1). This research has made use of the NASA/IPAC Extragalactic Database (NED) which is operated by the Jet Propulsion Laboratory, California Institute of Technology, under contract with the National Aeronautics and Space Administration. This work is based on observations collected at the European Organisation for Astronomical Research in the Southern Hemisphere under various ESO programme IDs, retrieved through the ESO Science Archive Facility. This publication makes use of data products from the Two Micron All Sky Survey, which is a joint project of the University of Massachusetts and the Infrared Processing and Analysis Center/California Institute of Technology, funded by the National Aeronautics and Space Administration and the National Science Foundation. This work made use of data from observations obtained as part of the VISTA Hemisphere Survey, ESO Progam, 179.A-2010 (PI: McMahon)






\bibliographystyle{mnras}
\bibliography{lens_search.bib} 



\appendix

\section{Example rejected emitter detection}
\label{app:rejem}

\begin{figure*}
	\centering
	\includegraphics[width=0.9\linewidth]{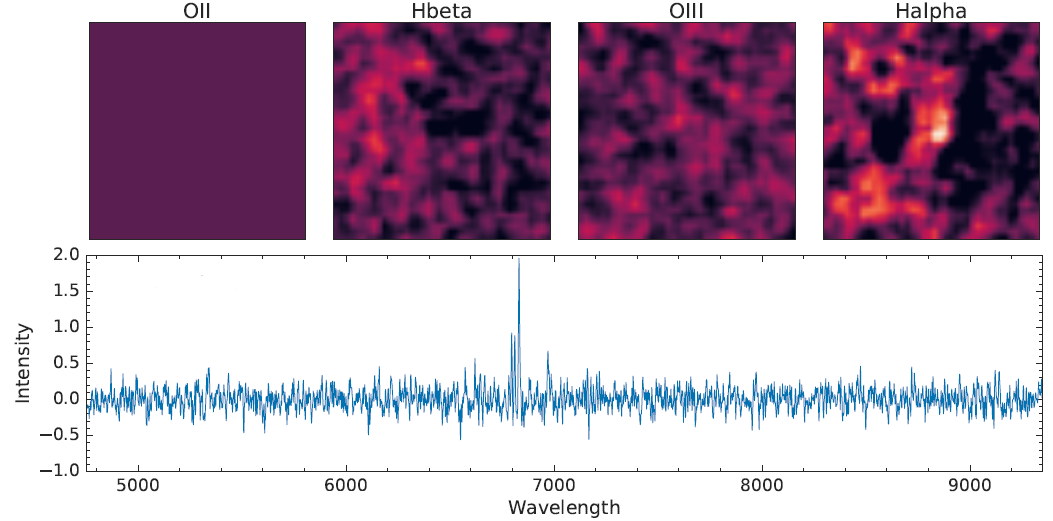}
	\caption{An example rejected detection from a MNELLS observation of J0058-1628. The four panels across the top show the spatial extent of the emission in narrow-band slices around the [{\sc O\,ii}], H\,$\beta$, [{\sc O\,iii}] and H\,$\alpha$, which tend to be the strongest optical emission lines. Here there is no [{\sc O\,ii}], as it out of the MUSE wavelength range for the fitted redshift. Below is the spectrum extracted within a 2\,arcsec diameter aperture, which shows that this is likely a residual of the lens subtraction. The best fit redshift is 0.03735.}
	\label{fig:exreject}
\end{figure*}

In this Appendix we show an example of a rejected detection (See Figure \ref{fig:exreject}). Here is it clear that the detection has found a residual in the H\,$\alpha$ at the redshift of the candidate galaxy. It can be automated to reject such emission during the detection process, using the known lens redshift. It is also clear that the [{\sc O\,ii}] emission line is not within the wavelength range of the MUSE instrument, and hence the panel which would contain a narrowband extraction is left blank.

\section{Multiple Close Emitters - More Information}
\label{app:MCE}

\subsection{2MASXJ00585131-1628092}
\label{J0058}

\begin{figure*}
	\centering
	\includegraphics[width=\linewidth]{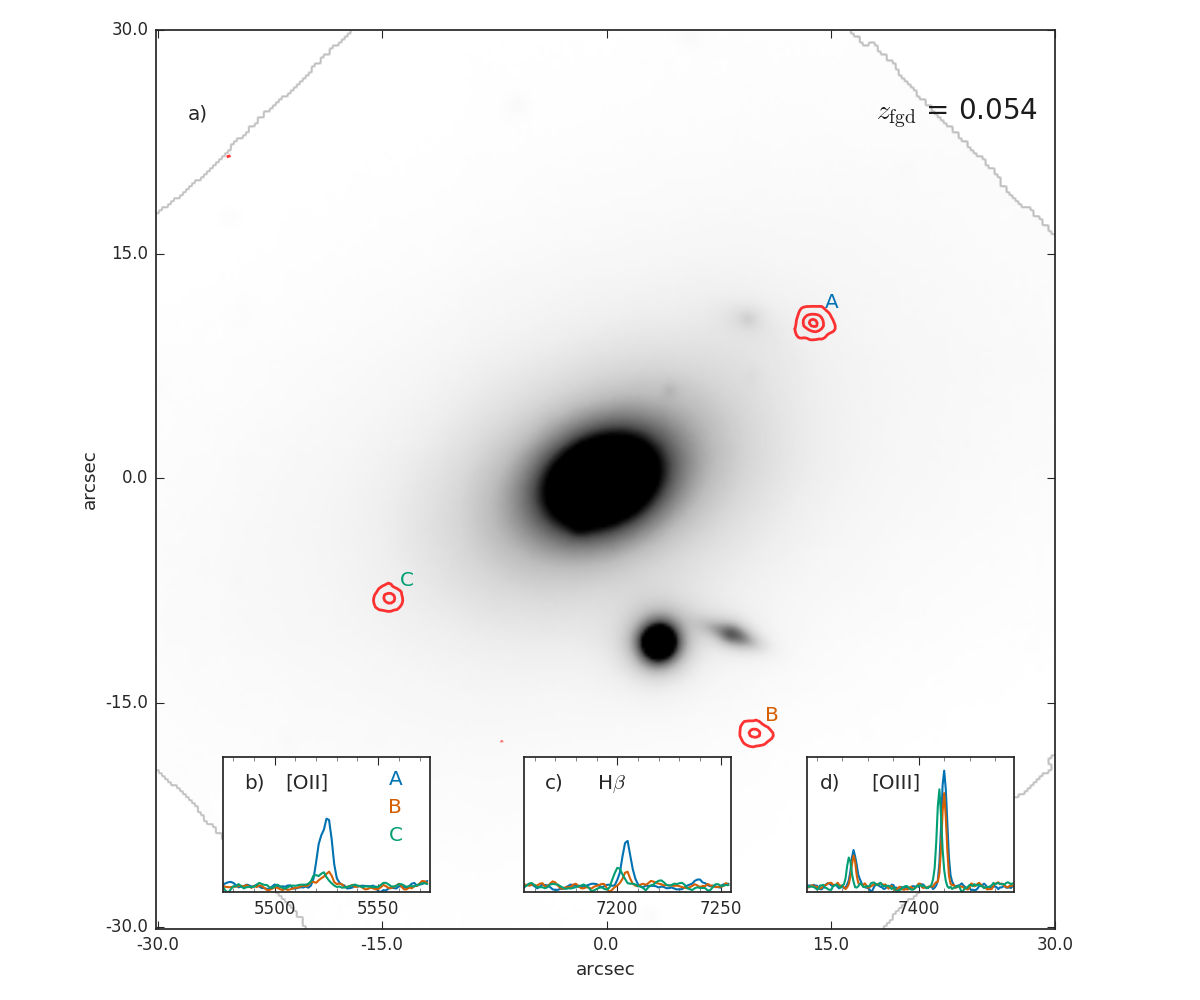}
	\caption{The MUSE data of J0058-1628, and the extracted background emitters. Panel a) The MUSE data of J0058-1628 collapsed over the full wavelength range, overlaid with contours of the background [{\sc O\,iii}] emission at common redshift, labelled A, B and C. Panels b,c,d) displays the [{\sc O\,ii}], H\,$\beta$ and [{\sc O\,iii}] emission for each of the three emitters. The line strength ratios vary between the images, with very weak [{\sc O\,ii}] emission only present in image A. Image C can be seen to have a velocity offset compared to A and B, in all three panels. Therefore we do not consider these to have originated from a single source.}
	\label{fig:J0058}
\end{figure*}

The first system we will discuss in this section is 2MASXJ00585131-1628092 (J0058-1628). This is a redshift 0.054 massive elliptical with a companion separated by 11.5\,arcsec. This system was previously observed with SINFONI, however, due to the small FoV of the instrument, three background clustered [{\sc O\,iii}] emitters were outside of the instrument FoV. However, in MUSE these emitters are detected, and shown in Figure \ref{fig:J0058}. 

The three emitters share very strong [{\sc O\,iii}] emission, but do not share the same [{\sc O\,ii}], and H\,$\beta$. There is also a significant velocity offset $\sim$\,200\,km\,s$^{-1}$ between C, and the other two, but only 10's\,km\,s$^{-1}$ between A and B, seen clearly in Figure \ref{fig:J0058}e). On face value, considering the images A, and B the system appears similar to J2357-3445, which we label a lens. In this case, the candidate images are separated by $\sim$28.28\,arcsec, and $>$\,15\,arcsec from the galaxy centre, which far exceeds the stellar galaxy-galaxy lensing regime. However, this lens does not appear to be part of a larger galaxy group, and hence is unlikely to have the large dark matter halo required to have such large image separation (Unlike J2357-3445). Finally, there is a weak H$\gamma$ line, seen in candidate image A, which is not seen in the spectrum of candidate B.

Therefore we label this system as a clustered group of background sources, and not a lensing system. Deeper observations and a more detailed understanding of the local matter distribution would be required to clearly define this system.

\subsection{2MASXJ13522523-3456007}
\label{J1352_app}

\begin{figure*}
	\centering
	\includegraphics[width=\linewidth]{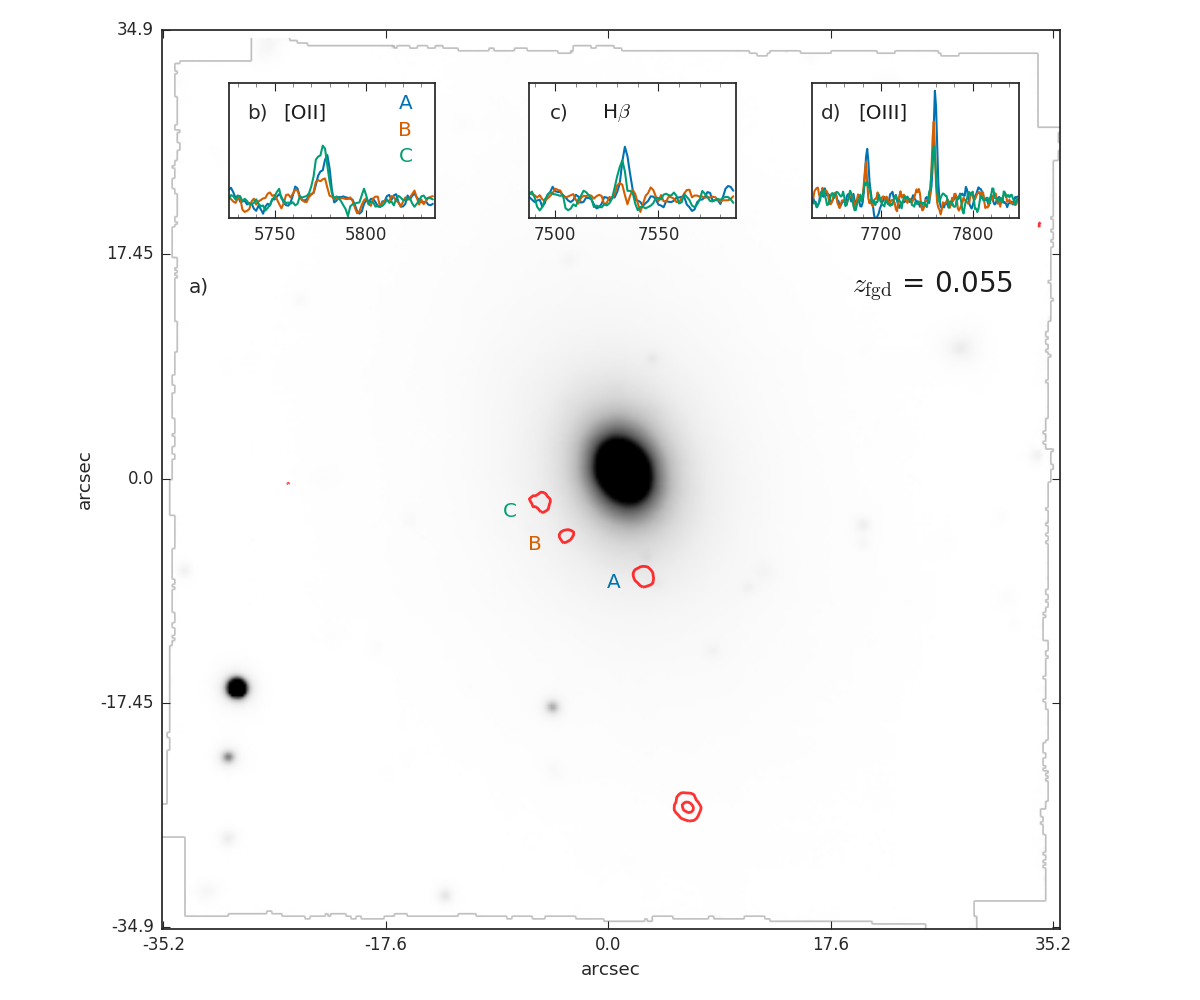}
	\caption{The MUSE data of J1352-3456, with the extracted background emitters spectra. Panel a) The MUSE imaging of J1352-3456 collapsed over the full wavelength range. Overlaid are the contours of the [{\sc O\,iii}] emission for the rejected candidate lensed images. Panels b,c,d) show the [{\sc O\,ii}], H\,$\beta$ and [{\sc O\,iii}] respectively for each of the labelled emitters. The line strength ratios vary, between images A and C, along with all three being offset in velocity space from each other (seen most clearly in panel c), H\,$\beta$).}
	\label{fig:J1352}
\end{figure*}

The second system is 2MASXJ1352252-3456007 (J1352-3456), an E/S0 galaxy at redshift 0.03817 \citep{Jones2009}, with no nearby galaxies of comparable size. Our MNELLS observations detect a group of three emitters separated from the galaxy centre by between 5 and 7\,arcsec. These are detected at redshift $\sim$\,0.549, via strong [{\sc O\,iii}] and weaker [{\sc O\,ii}]. A fourth emitter can be seen in Figure \ref{fig:J1352}a) to the south, but is poorly configured with the other three sources, and has a velocity offset of 210\,km\,s$^{-1}$ so is not considered part of the candidate lens system.

The spectra of the three objects are very similar, with A and C of comparable line strengths, and B the faintest. The largest velocity offset is between A and C is $\sim$\,90\,km\,s$^{-1}$, with B and C offset by $<$\,50\,km\,s$^{-1}$. However, in this system the problematic feature is the distance from the lens to the images, and the separation between the images. Where these images close, and hint at a some linked structure this may bear a resemblance to the lens reported in \citet{Smith2017b}, with a large arc, and no clearly observable counter-image. However, no evidence is present to define this as an arc. The separation from the lens of $\sim$\,6.5\,arcsec lies at the limit of twice the expected \rein, for a similarly massive galaxy and hence unlikely to lie within the strongly lensing regime. Secondly to form three distinct images, each separated by over 1\,arcsec is an unlikely lensing configuration. In addition there is no evidence for these candidate images forming a single arc, as no emission is detected in the spaces between C, B and A.

In order to understand this lensing system much deeper data would be required to test for the presence of a faint inner image, very close to the candidate lens core, or to identify C, B and A as a single arc. Furthermore these potential images lie at a distance which would only add weak constraints to the required mass for there to be no detectable counter image. Hence, we label these emitters a group of background galaxies.

%
%

%
%
%
\end{document}